\documentclass[twocolumn,showpacs,preprintnumbers,amsmath,amssymb,nofootinbib]{revtex4-1}

\usepackage{graphicx}
\usepackage{dcolumn}
\usepackage{bm}
\usepackage{amsfonts}

\def\bequ{\begin{equation}}
\def\eequ{\end{equation}}
\def\be{\begin{equation}}
\def\ee{\end{equation}}

\begin{document}
\title{Charged Dirac perturbations on Reissner-Nordstr\"om-Anti-de Sitter spacetimes: \\quasinormal modes with Robin boundary conditions
}

\author{Mengjie Wang$^1$}
\email{mjwang@hunnu.edu.cn, corresponding author.}
\author{Carlos Herdeiro$^2$}
\email{herdeiro@ua.pt}
\author{Jiliang Jing$^1$}
\email{jljing@hunnu.edu.cn}
\affiliation{\vspace{2mm}
$^1$Department of Physics, Key Laboratory of Low Dimensional Quantum Structures and
Quantum Control of Ministry of Education, and Synergetic Innovation Center for Quantum Effects and Applications, Hunan Normal University,
Changsha, Hunan 410081, P.R. China
\vspace{1.3mm}
\\
$^2$Departamento de Matem\'atica da Universidade de Aveiro and CIDMA, Campus de Santiago, 3810-183 Aveiro, Portugal \vspace{1mm}}%

\date{October 2019}

\begin{abstract}
We study charged Dirac quasinormal modes (QNMs) on Reissner-Nordstr\"om-Anti-de Sitter (RN-AdS) black holes with generic Robin boundary conditions, by extending our earlier work of neutral Dirac QNMs on Schwarzschild-AdS black holes. We first derive the equations of motion for charged Dirac fields on a RN-AdS background. To solve these equations we impose a requirement on the Dirac field: that its energy flux should vanish at asymptotic infinity. A set of \textit{two} Robin boundary conditions compatible with QNMs is consequently found. By employing both analytic and numeric methods, we then obtain the quasinormal spectrum for charged Dirac fields, and analyse the impact of various parameters, in particular of electric charge. An analytic calculation shows \textit{explicitly} that the charge coupling between the black hole and the Dirac field does \textit{not} trigger superradiant instabilities. Numeric calculations, on the other hand, show quantiatively that Dirac QNMs may change substantially due to the electric charge. Our results illustrate how vanishing energy flux boundary conditions, as a generic principle, are applicable not only to neutral but also to electrically charged fields.
\end{abstract}

\maketitle

\section{Introduction}
Black hole perturbation theory is a useful approach to study interactions between black holes (BHs) and fundamental fields. It plays a vital role in various contexts, ranging from gravitational wave physics to fundamental particle physics~\cite{Barack:2018yly}. At the linear level, the BH-fields system is, \textit{a priori},  described by a set of partial differential equations. However, the celebrated work of Regge and Wheeler~\cite{Regge:1957td}, Zerilli~\cite{Zerilli:1970se} and Teukolsky~\cite{Teukolsky:1973ha}, showed that perturbation equations of various massless spin fields on the paradigmatic Schwarzschild and Kerr black holes of General Relativity can be formulated in terms of simpler ordinary differential equations.

Such perturbation equations set the foundation to explore properties of the BH-fields linear interactions. A key issue is then the choice of boundary conditions. In asymptotically flat spacetimes, one may study \textit{scattering states} by imposing both ingoing and outgoing wave boundary conditions at infinity, \textit{quasi-bound states} by imposing decaying boundary condition at infinity, and \textit{quasinormal modes} (QNMs) by imposing outgoing wave boundary condition at infinity. In asymptotically anti-de Sitter (AdS) spacetimes, the study of QNMs, in particular, has attracted a lot of attention in the last two decades due to the AdS/CFT correspondence~\cite{Maldacena:1997re} - see for example~\cite{Kokkotas:1999bd,Berti:2009kk,Konoplya:2011qq} and references therein.

Asymptotically AdS spacetimes have a timelike conformal boundary. How to define QNMs in such spacetimes is a partly open question. The investigation of QNMs in asymptotically AdS spacetimes was initiated considering scalar fields on Schwarzschild-AdS BHs~\cite{Horowitz:1999jd,Chan:1996yk,Chan:1999sc}, where the boundary condition was set as the vanishing of the scalar field itself at the asymptotic boundary. Subsequently, generalizations to other spin fields (Maxwell/gravitational/Dirac/Rarita-Schwinger fields, etc.), used the same scalar-like boundary condition~\cite{Cardoso:2001bb,Cardoso:2001hn,Cardoso:2001vs,Cardoso:2003cj,Berti:2003ud,Giammatteo:2004wp,Jing:2005ux,Jing:2005uy,Miranda:2005qx,Zhang:2006bc,Liu:2008ds,Berti:2009wx,Cai:2010tr,Li:2010sv}. The drawback of such boundary condition is that it cannot be used for Teukolsky variables, as shown explicitly for Maxwell fields~\cite{PhysRevD.92.124006}, so it is not a generic boundary condition.

Some physical guidance is useful in understanding more generic boundary conditions. The AdS boundary plays the role of a perfectly reflecting mirror. Based on this idea, a generic principle for the boundary condition of an arbitrary spin field was proposed: that the energy flux of the field should vanish at the asymptotic boundary~\cite{PhysRevD.92.124006}\footnote{Vanishing energy flux also leads to vanishing angular momentum flux, as shown explicitly for the Maxwell field in~\cite{Wang:2016dek}.}. This principle has already been applied to the Maxwell field, and led to two different quantitative boundary conditions, dubbed as vanishing energy flux boundary conditions, and consequently to two distinct sets of QNMs~\cite{PhysRevD.92.124006,Wang:2015fgp,Wang:2016dek,Wang:2016zci}.

In order to verify the universal applicability of the vanishing energy flux boundary conditions, we have initiated a systematic study of Dirac field perturbations on asymptotically AdS spacetimes~\cite{Wang:2017fie}. As a follow up on this work, herein we generalize the previous study of neutral Dirac fields~\cite{Wang:2017fie} by adding charge to both the background and the field. This will illustrate that the vanishing energy flux boundary conditions may be applied not only to neutral but also to electrically charged fields.

The structure of this paper is organized as follows. In Section~\ref{seceq} we briefly introduce the Reissner-Nordstr\"om-AdS ( RN-AdS) geometry and derive the corresponding massless charged Dirac equations in the $\gamma$ matrices formalism. In Section~\ref{secbc} we derive equations of motion and construct the energy flux for charged Dirac fields on  RN-AdS BHs. By requiring the energy flux to vanish at the boundary, one obtains two sets of Robin boundary conditions. In Section~\ref{secana} we solve the charged Dirac equations \textit{analytically} in the small BH and low frequency approximations, by using a standard matching method. We obtain the imaginary part of the charged Dirac QNMs and prove \textit{explicitly} that superradiant instabilities \textit{do not} exist. Three different numerical methods and various results are presented in Section~\ref{secnum}, to illustrate the effect of the electric charge (both of the field and of the background) on the two sets of QNMs, varying the BH size $r_+$, the angular momentum quantum number $\ell$, and the overtone number $N$. Final remarks and conclusions are presented in the last section. Some formulas of the Horowitz-Hubeny method are left to the Appendix.

We use natural units and a $(+ - - -)$ signature throughout.

\section{background geometry and field equations}
\label{seceq}
In this section, we briefly review the four-dimensional RN-AdS BH, and derive the massless charged Dirac equations on this background.

\subsection{Reissner-Nordstr\"om-AdS BHs}
The line element of a four-dimensional RN-AdS BH may be written as
\begin{equation}
ds^2=\dfrac{\Delta_r}{r^2}dt^2-\dfrac{r^2}{\Delta_r}dr^2-r^2d\theta^2-r^2\sin^2\theta d\varphi^2 \;,\label{metric}
\end{equation}
with the metric function
\begin{eqnarray}
\Delta_r\equiv r^2\left(1+\frac{r^2}{L^2}\right)-2Mr+Q^2\;,\label{metricfunc}
\end{eqnarray}
where $L$ is the AdS radius, $M$ and $Q$ are the mass and electric charge parameters of the background.
The mass parameter $M$ can be expressed as
\begin{equation}
M=\dfrac{r_+^2(L^2+r_+^2)+Q^2L^2}{2r_+L^2}\;,\nonumber
\end{equation}
where the outer (event) horizon , at $r=r_+$, is determined as the largest root of $\Delta_r(r_+)$=0.
The Hawking temperature is then given by
\begin{equation}
T_H=\dfrac{\kappa}{2\pi}=\dfrac{1}{4\pi r_+^3}\left[r_+^2\left(1+3\dfrac{r_+^2}{L^2}\right)-Q^2\right]\;.\label{hawkingT}
\end{equation}
For non-extremal BHs, the background charge $Q$ is constrained by a critical charge $Q_c$
\begin{equation}
Q< Q_c\equiv r_+\sqrt{1+3\dfrac{r_+^2}{L^2}}\;,\label{criticalQ}
\end{equation}
where $Q_c$ is the background charge for extremal BHs.
The electromagnetic potential of the RN-AdS BH is
\begin{equation}
A=\left(-\dfrac{Q}{r}+C\right)dt\;,\label{potential}
\end{equation}
where the constant $C$ is a gauge choice. The choice of this constant, as we will see in the following, modifies the equations of motion and the boundary conditions, by simply shifting the frequency $\omega$ to $\omega+qC$, where $q$ is the field charge.

\subsection{Charged Dirac equations}

The equations of motion for a massless Dirac field may be derived by the $\gamma$ matrices formalism~\cite{Unruh:1973bda}. A charged massless Dirac field obeys the equation
\begin{equation}
\gamma^\mu(\mathcal{D}_\mu-\Gamma_\mu)\Psi=0\;,\label{ChargedDiraceqU1}
\end{equation}
with $\mathcal{D}_\mu\equiv \partial_\mu-iqA_\mu$, where $q$ is the field charge, and the electromagnetic potential $A_\mu$ is given by Eq.~\eqref{potential}. The $\gamma$ matrices are defined as~\cite{Wang:2017fie}
\begin{align}
&\gamma^t=\sqrt{\frac{r^2}{\Delta_r}}\gamma^0\;,\;\;\;\;\;\;\gamma^r=\sqrt{\frac{\Delta_r}{r^2}}\gamma^3\;,\nonumber\\
&\gamma^\theta=\frac{1}{r}\gamma^1\;,\;\;\;\;\;\;\;\;\;\;\;\;\gamma^\varphi=\frac{1}{r\sin\theta}\gamma^2\;,\label{gammams}
\end{align}
with the ordinary flat spacetime Dirac matrices $\gamma^i (i=0, 1, 2, 3)$ being provided in the Bjorken-Drell representation~\cite{Bjorken:100769}. The spin connection is
\begin{equation}
\Gamma_\mu=-\frac{1}{8}(\gamma^a\gamma^b-\gamma^b\gamma^a)\Sigma_{ab\mu}\;,\label{spincon}
\end{equation}
where
\begin{equation}
\Sigma_{ab\mu}=e_a^\nu(\partial_\mu e_{b\nu}-\Gamma^\alpha_{\nu\mu}e_{b\alpha})\  ,\nonumber
\end{equation}
in terms of the tetrad $e_a^\mu$.
Eq.~\eqref{ChargedDiraceqU1} leads to a set of first order differential equations, which are coupled but with separated variables,
\begin{align}
&\Delta_r^{1/2}\left(\dfrac{d}{dr}-\dfrac{iK_r}{\Delta_r}\right)R_1(r)=\lambda R_2(r)\;,\label{Chargedfirstorderr1}\\
&\Delta_r^{1/2}\left(\dfrac{d}{dr}+\dfrac{iK_r}{\Delta_r}\right)R_2(r)=\lambda R_1(r)\;,\label{Chargedfirstorderr2}\\
&\left(\dfrac{d}{d\theta}-\dfrac{m}{\sin\theta}\right)S_1(\theta)=\lambda S_2(\theta)\;,\label{Chargedfirstorderang1}\\
&\left(\dfrac{d}{d\theta}+\dfrac{m}{\sin\theta}\right)S_2(\theta)=-\lambda S_1(\theta)\;,\label{Chargedfirstorderang2}
\end{align}
with $K_r \equiv (\omega+qC) r^2-qQr$, by using the Dirac field ansatz
\begin{equation}
\Psi=\left(
\begin{matrix}
\eta\\
\eta
\end{matrix}
\right)\;,\;\eta=\frac{e^{-i\omega t}e^{im\varphi}}{(\Delta_rr^2\sin^2\theta)^{1/4}}
\left(
\begin{matrix}
R_1(r)S_1(\theta)\\
R_2(r)S_2(\theta)
\end{matrix}
\right)\;,\label{Chargedfielddecom}
\end{equation}
where $\omega$ and $m$ are the frequency and azimuthal number of Dirac fields, respectively.

For the convenience in the following analytic and numeric calculations, we transform the above first order equations to the second order ones. It is straightforward to obtain the radial equations
\begin{align}
&\Delta_r^{1/2}\dfrac{d}{dr}\left(\Delta_r^{1/2}\dfrac{d}{dr}\right)R_{1}(r)+H_1(r)R_{1}(r)=0\;,\label{ChargedDiracU2radialR1}\\
&\Delta_r^{1/2}\dfrac{d}{dr}\left(\Delta_r^{1/2}\dfrac{d}{dr}\right)R_{2}(r)+H_2(r)R_{2}(r)=0\;,\label{ChargedDiracU2radialR2}
\end{align}
with
\begin{align}
&H_1(r)=\dfrac{K_r^2+\tfrac{i}{2}K_r\Delta_r^\prime}{\Delta_r}-2i(\omega+qC) r+iqQ-\lambda^2\;,\nonumber\\
&H_2(r)=\dfrac{K_r^2-\tfrac{i}{2}K_r\Delta_r^\prime}{\Delta_r}+2i(\omega+qC) r-iqQ-\lambda^2\;,\nonumber
\end{align}
where $\prime$ represents a derivative with respect to $r$, and the angular equations
\begin{align}
&\dfrac{d^2S_{1}(\theta)}{d\theta^2}+\left(-\dfrac{m^2}{\sin^2\theta}+m\dfrac{\cos\theta}{\sin^2\theta}+\lambda^2\right)S_{1}(\theta)=0\;,\label{ChargedDiracU2angularS1}\\
&\dfrac{d^2S_{2}(\theta)}{d\theta^2}+\left(-\dfrac{m^2}{\sin^2\theta}-m\dfrac{\cos\theta}{\sin^2\theta}+\lambda^2\right)S_{2}(\theta)=0\;,\label{ChargedDiracU2angularS2}
\end{align}
where $S_1(\theta)$ and $S_2(\theta)$ are spin-weighted spherical harmonics with the corresponding eigenvalue $\lambda^2=(\ell+\frac{1}{2})^2$~\cite{Dolan:2009kj}.

The radial part of the second order differential equations, Eqs.~\eqref{ChargedDiracU2radialR1}--Eq.~\eqref{ChargedDiracU2radialR2}, will be the main focus of interest in the remaining sections.

\section{boundary conditions}
\label{secbc}
In order to obtain quasinormal frequencies for charged Dirac fields on RN-AdS BHs, we have to solve the differential equations \eqref{ChargedDiracU2radialR1}--\eqref{ChargedDiracU2radialR2} with \textit{proper} boundary conditions. At the horizon, we impose purely ingoing boundary conditions, as usual. At the asymptotic region, we follow the generic principle proposed in~\cite{PhysRevD.92.124006,Wang:2015fgp,Wang:2016dek,Wang:2016zci} by requiring that the \textit{energy flux} vanishes\footnote{A scalar-like boundary condition has been previously employed to study QNMs for neutral Dirac fields on RN-AdS BHs~\cite{Giammatteo:2004wp,Jing:2005ux,Jing:2005uy}.}.

To implement this principle, we start with the energy-momentum tensor for charged Dirac fields
\begin{equation}
T_{\mu \nu}=\dfrac{i}{8\pi}\bar{\Psi}\left[\gamma_\mu(\mathcal{D}_\nu-\Gamma_\nu)+\gamma_\nu(\mathcal{D}_\mu-\Gamma_\mu)\right]\Psi+c.c. \;,\label{EMTensorDirac}
\end{equation}
where $\bar{\Psi}\equiv \Psi^\dag\gamma^0$, and $c.c.$ stands for complex conjugate of the preceding terms. Note that $\gamma_\mu=g_{\mu\nu}\gamma^\nu$, where $\gamma^\nu$ is given in Eq.~\eqref{gammams}, the spin connection $\Gamma_\mu$ is given in Eq.~\eqref{spincon}, and $\Psi^\dag$ is the hermitian conjugate of $\Psi$.

The energy flux through a 2-sphere at radial coordinate $r$ is
\begin{equation}
\mathcal{F}|_r=\int_{S^2} \sin\theta d\theta d\varphi\; r^2 T^r_{\;\;t}\;, \label{Chargedflux1}
\end{equation}
where
\begin{equation}
T^r_{\;\;t}=T^r_{\;\;t,\;\uppercase\expandafter{\romannumeral1}}+T^r_{\;\;t,\;\uppercase\expandafter{\romannumeral2}}\;,\nonumber\\
\end{equation}
and
\begin{equation}
T^r_{\;\;t,\;\uppercase\expandafter{\romannumeral1}}=\dfrac{\omega+\omega^{\ast}+2qA_t}{2\pi r^2\sin\theta}\left(|R_1|^2|S_1|^2-|R_2|^2|S_2|^2\right)\ .
\end{equation}
Here $\omega^{\ast}$ is the complex conjugate of $\omega$ and $T^r_{\;\;t,\;\uppercase\expandafter{\romannumeral2}}$  vanishes after integrating over the sphere. Then, the energy flux becomes
\begin{equation}
\mathcal{F}|_r\propto(\omega+\omega^{\ast}+2qA_t)\left(|R_1|^2-|R_2|^2\right)\;,\label{Chargedflux2}
\end{equation}
up to a factor independent of the radial coordinate $r$, and where the angular functions $S_{1,2}(\theta)$ have been normalized
\begin{equation}
\int_0^\pi d\theta\; |S_{1,2}(\theta)|^2=1\;.\nonumber
\end{equation}

To obtain the explicit expressions of boundary condition for $R_{1}$, we make the asymptotic expansion from Eq.~\eqref{ChargedDiracU2radialR1}, and get
\begin{equation}
R_1 \sim \alpha_1+\beta_1\dfrac{L}{r}+\mathcal{O}\left(\dfrac{L^2}{r^2}\right)\;,\label{ChargedasysolR1}
\end{equation}
where $\alpha_1$ and $\beta_1$ are two integration constants.

With the relation between $R_1$ and $R_2$ in Eq.~\eqref{Chargedfirstorderr1}, and making use of expansion for $R_1$ in Eq.~\eqref{ChargedasysolR1}, at infinity the energy flux in Eq.~\eqref{Chargedflux2} becomes
\begin{eqnarray}
\mathcal{F}|_{r,\infty}\propto |\alpha_1|^2-\dfrac{1}{\lambda^2}|i(\omega+qC) L\alpha_1+\beta_1|^2\;.\label{Chargedfluxinf}
\end{eqnarray}
Now we are able to impose the \textit{energy flux vanishing boundary conditions}, i.e. $\mathcal{F}|_{r,\infty}=0$, which implies
\begin{equation}
\lambda^2|\alpha_1|^2-|i(\omega+qC) L\alpha_1+\beta_1|^2=0\;.
\end{equation}
It is easy to solve this quadratic equation and obtain the two solutions\footnote{The relative phase between two moduli has been fixed by calculating normal modes.}
\begin{align}
\dfrac{\alpha_1}{\beta_1}=\dfrac{-i}{\ell+\frac{1}{2}-(\omega+qC) L}\;,\label{Chargedbc1}\\
\dfrac{\alpha_1}{\beta_1}=\dfrac{i}{\ell+\frac{1}{2}+(\omega+qC) L}\;,\label{Chargedbc2}
\end{align}
which are exactly the same conditions obtained for the neutral Dirac fields, when $q=0$. These two boundary conditions imply two branches of QNMs for charged Dirac fields, following the same logic as for the neutral Dirac and Maxwell cases~~\cite{Wang:2017fie,PhysRevD.92.124006,Wang:2015fgp,Wang:2016dek,Wang:2016zci}.

The boundary conditions for $R_2$ may be also obtained
\begin{equation}
\dfrac{\alpha_2}{\beta_2}=\dfrac{i}{\ell+\frac{1}{2}-(\omega+qC) L}\;,\;\;\;\dfrac{\alpha_2}{\beta_2}=\dfrac{-i}{\ell+\frac{1}{2}+(\omega+qC) L}\;,\label{Chargedbcanother}
\end{equation}
by repeating the same procedure as illustrated above, and where $\alpha_2$ and $\beta_2$ are the first two expansion coefficients for $R_2$ at infinity.

As one may check, solving the radial equation~\eqref{ChargedDiracU2radialR1} with the corresponding boundary conditions~\eqref{Chargedbc1},~\eqref{Chargedbc2} and the radial equation~\eqref{ChargedDiracU2radialR2} with the corresponding boundary conditions~\eqref{Chargedbcanother}, the same quasinormal frequencies could be obtained. Therefore, for concreteness and without loss of generality, in the following we only focus on the $R_1$ equation and the corresponding boundary conditions.

\section{Analytic calculations}
\label{secana}
In this section we calculate the imaginary part of QNMs for the charged Dirac equation~\eqref{ChargedDiracU2radialR1} \textit{analytically}, under the two Robin boundary conditions given in Eqs.~\eqref{Chargedbc1}--\eqref{Chargedbc2}, by using the \textit{asymptotic matching method}. Since the gauge constant $C$ only changes the real part of the Dirac QNMs, we can safely take $C=0$ without loss of generality. Our goal is to show \textit{explicitly} that charged Dirac fields on RN-AdS BHs do \textit{not} trigger superradiant instabilities.

In order to employ the asymptotic matching method, we shall first divide the exterior region to the event horizon into the \textit{near region} ($r-r_+\ll1/\omega$) and the \textit{far region} ($r-r_+\gg r_+$). Eq.~\eqref{ChargedDiracU2radialR1} may then be solved separately in these two regions in the small charge coupling limit ($qQ\ll1$). In the low frequency approximation ($\omega r_+\ll1$), the solutions obtained in those two regions are both valid in an \textit{overlap region} ($r_+\ll r-r_+\ll1/\omega$). Then, imposing physically relevant boundary conditions, the QNMs can be computed. We further require that the RN-AdS BHs are small ($r_+\ll L$), so that one may obtain quasinormal frequencies perturbatively, as deformations of the empty AdS normal modes.

\subsection{Near region solution}
To solve Eq.~\eqref{ChargedDiracU2radialR1} in the near region, it is convenient to define a new dimensionless variable
\begin{equation}
z\equiv \dfrac{r-r_+}{r-r_-}\;,\nonumber
\end{equation}
where $r_+$ and $r_-$ are the event horizon and the Cauchy horizon radial coordinate of RN-AdS BH, respectively. This new variable together with the small BH approximation ($r_+\ll L$) bring Eq.~\eqref{ChargedDiracU2radialR1} into
\begin{equation}
z(1-z)\dfrac{d^2R_1}{dz^2}+\dfrac{1-3z}{2}\dfrac{dR_1}{dz}+\left(\hat{\omega}\dfrac{1-z}{z}-\dfrac{\bar{\lambda}^2}{1-z}\right)R_1=0 \;,\label{chargedneareq1}
\end{equation}
where
\begin{equation}
\hat{\omega}\equiv \left(\bar{\omega}+\dfrac{i}{4}\right)^2+\dfrac{1}{16}\;,\;\;\;\bar{\omega}\equiv \left(\omega-\dfrac{qQ}{r_+}\right)\dfrac{r_+^2}{r_+-r_-}\;,\nonumber
\end{equation}
and
\begin{equation}
\bar{\lambda}^2=\left(\bar{\ell}+\dfrac{1}{2}\right)^2\;,\;\;\;\bar{\ell}\equiv \ell+\epsilon\;.\nonumber
\end{equation}
Here, $\epsilon$ is a small quantity of order $r_+/L$. The small correction introduced by $\epsilon$ implies that $\bar{\ell}$ is \textit{not} an exact half integer anymore. Observe that $\bar{\omega}<0$ when $\omega r_+<qQ$; the final QNMs expression, however, will not admit a sign change for the imaginary part.

By imposing the ingoing wave boundary condition at the event horizon, the solution for Eq.~\eqref{chargedneareq1} is
\begin{equation}
R_1\sim z^{\frac{1}{2}-i\bar{\omega}}(1-z)^{\ell+\frac{1}{2}}\;F(a,b,c;z)\;,\label{chargednearsol}
\end{equation}
where $F(a,b,c;z)$ is the hypergeometric function, with
\begin{equation}
a=\bar{\ell}+1\;,\;\;\;
b=\bar{\ell}+\frac{3}{2}-2i\bar{\omega}\;,\;\;\;
c=\frac{3}{2}-2i\bar{\omega}\;.\nonumber
\end{equation}
In order to match with the far region solution given in the next subsection, we shall expand the solution given in Eq.~\eqref{chargednearsol}, at large $r$. By taking the limit $z\rightarrow1$ and using the properties of the hypergeometric function~\cite{abramowitz+stegun}, we then obtain
\begin{equation}
R_1 \;\sim \; \Gamma(c)\left[\dfrac{R^{\rm near}_{1,1/r}}{r^{\ell+\frac{1}{2}}} +R^{\rm near}_{1,r} r^{\ell+\frac{1}{2}}\right]
\;,\label{chargednearsolfar}
\end{equation}
where
\begin{align}
R^{\rm near}_{1,1/r} & \equiv \dfrac{\Gamma(-2\bar{\ell}-1) (r_+-r_-)^{\ell+\frac{1}{2}}}{\Gamma(-\bar{\ell})\Gamma(\frac{1}{2}-\bar{\ell}-2i\bar{\omega})} \ ,\nonumber \\
R^{\rm near}_{1,r} &\equiv \dfrac{\Gamma(2\bar{\ell}+1)(r_+-r_-)^{-\ell-\frac{1}{2}}}{\Gamma(\bar{\ell}+1)\Gamma(\bar{\ell}+\frac{3}{2}-2i\bar{\omega})} \ .
\end{align}

\subsection{Far region solution}
In the far region the BH may be neglected ($M\rightarrow0, Q\rightarrow0$). Then, the solution for Eq.~\eqref{ChargedDiracU2radialR1} is the same as for neutral Dirac fields~\cite{Wang:2017fie}. To match with the near region solution, one has to expand the far region solution at small $r$, which is given by~\cite{Wang:2017fie}
\begin{equation}
R_1\;\sim\;\dfrac{R^{\rm far}_{1,1/r}}{r^{\ell+\frac{1}{2}}}+R^{\rm far}_{1,r}r^{\ell+\frac{1}{2}}\;,\label{chargedfarsolnear}
\end{equation}
with
\begin{eqnarray}
&&R^{\rm far}_{1,1/r}  \equiv 2^{-2\ell-1} (iL)^{2\ell+1} C_1 \ ,\nonumber \\
&&R^{\rm far}_{1,r} \equiv  C_2\;.\nonumber
\end{eqnarray}
The two integration constants $C_1$ and $C_2$ are determined by the boundary conditions~\cite{Wang:2017fie}, i.e.
\begin{equation}
\dfrac{C_1}{C_2}=2^{2\ell+1}\dfrac{\ell}{\ell-\omega L}\dfrac{\mathcal{A}_1}{\mathcal{A}_2}\;,\label{c1c2rel1}
\end{equation}
corresponding to the first boundary condition given in Eq.~\eqref{Chargedbc1}, and
\begin{equation}
\dfrac{C_1}{C_2}=2^{2\ell+1}\ell\dfrac{\ell+1+\omega L}{\ell+1}\dfrac{\mathcal{A}_3}{\mathcal{A}_4}\;,\label{c1c2rel2}
\end{equation}
corresponding to the second boundary condition given in Eq.~\eqref{Chargedbc2}, where
\begin{align}
\mathcal{A}_1=\;&F\left(\ell+\frac{1}{2},\ell+1+\omega L,2\ell+2;2\right)\nonumber\\&+F\left(\ell+\frac{3}{2},\ell+1+\omega L,2\ell+2;2\right)\;,\nonumber\\
\mathcal{A}_2=\;&F\left(\frac{1}{2}-\ell,1-\ell+\omega L,1-2\ell;2\right)\;,\nonumber\\
\mathcal{A}_3=\;&F\left(\frac{3}{2}+\ell,2+\ell+\omega L,2\ell+3;2\right)\;,\nonumber\\
\mathcal{A}_4=\;&2\ell F\left(-\ell-\frac{1}{2},-\ell+\omega L,-2\ell;2\right)\nonumber\\&+(\ell-\omega L) F\left(-\ell+\frac{1}{2},-\ell+1+\omega L,1-2\ell;2\right)\;.\nonumber
\end{align}

\subsection{Overlap region}
In the low frequency approximation ($\omega r_+\ll1$), the near region solution given in Eq.~\eqref{chargednearsolfar} and the far region solution given in Eq.~\eqref{chargedfarsolnear} are both valid in an overlap region. By imposing the matching condition $R^{\rm near}_{1,r}R^{\rm far}_{1,1/r}=R^{\rm far}_{1,r}R^{\rm near}_{1,1/r}$, we obtain
\begin{align}
&\dfrac{\Gamma(\bar{\ell}+1)}{\Gamma(2\bar{\ell}+1)}\dfrac{\Gamma(\bar{\ell}+\frac{3}{2}-2i\bar{\omega})}{\Gamma(-\bar{\ell}+\frac{1}{2}-2i\bar{\omega})}\dfrac{\Gamma(-2\bar{\ell}-1)}{\Gamma(-\bar{\ell})}\left(\dfrac{r_+-r_-}{L}\right)^{2\ell+1}\nonumber\\
&=\left(\dfrac{i}{2}\right)^{2\ell+1}\dfrac{C_1}{C_2}\;.\label{chargedmatching}
\end{align}
In empty AdS, the left term in the above equation vanishes, so that we have to require $C_1=0$. This condition leads to AdS normal modes. From Eqs.~\eqref{c1c2rel1} and~\eqref{c1c2rel2}, together with the condition $C_1=0$, one may get two sets of normal modes~\cite{Wang:2017fie}
\begin{equation}
\omega_{1,N}L=2N+\ell+1\;,\;\;\;\omega_{2,N}L=2N+\ell+2\;,\label{normalmodes}
\end{equation}
where $\omega_{1,N}$ and $\omega_{2,N}$ refer to the frequencies corresponding to the first and second boundary conditions with the overtone number $N=0,1,2,\cdot\cdot\cdot$, and the angular momentum quantum number $\ell=1/2,3/2,\cdot\cdot\cdot$. %

By taking into account of the existence of a (small) BH, a correction to the frequency (which may be complex) is introduced
\begin{equation}
\omega_j L=\omega_{j,N} L+i\delta_j\;,\label{modeperturb}
\end{equation}
where $\omega_{j,N}$ are the normal modes given by Eq.~\eqref{normalmodes}, and $j=1,2$ correspond to the two different boundary conditions. Here the real part of $\delta$ describes the damping of the AdS normal modes which become QNMs. Expanding $\omega$ in \mbox{Eq.~\eqref{chargedmatching}} on top of empty AdS, by using Eq.~\eqref{modeperturb}, one obtains
\begin{align}
&\mbox{Re}\delta_j=-\bar{\omega}\sigma_j\dfrac{\Gamma^2(\ell+1)}{\Gamma(2\ell+1)\Gamma(2\ell+2)}\prod^{\ell-1/2}_{p=1}(p^2+4\bar{\omega}^2)\nonumber\\&
\left((\ell+\dfrac{1}{2})\coth2\pi\bar{\omega}+\dfrac{2\bar{\omega}}{\pi\epsilon}\right)\left(\dfrac{r_+-r_-}{L}\right)^{2\ell+1}\;,\label{generaldelta}
\end{align}
with
\[\sigma_j=\left\{\begin{array}{ll}
(-1)^{\ell+1/2}\dfrac{i(2N+1)}{\ell}\dfrac{\mathcal{A}_2^\prime}{\mathcal{A}_1^\prime}&\;\;\;\text{$j=1$}\;,\\
(-1)^{\ell-1/2}\dfrac{i(\ell+1)}{\ell(2N+2\ell+3)}\dfrac{\mathcal{A}_4^\prime}{\mathcal{A}_3^\prime}&
\;\;\;\text{$j=2$}\;,
\end{array}\right.\]
which are real and positive, and where
\begin{align}
\mathcal{A}_1^\prime&=F^{(0,1,0,0)}\left(\frac{1}{2}+\ell,2N+2\ell+2,2+2\ell;2\right)\nonumber\\&+F^{(0,1,0,0)}\left(\frac{3}{2}+\ell,2N+2\ell+2,2+2\ell;2\right)\;,\nonumber\\
\mathcal{A}_2^\prime&=F\left(\frac{1}{2}-\ell,2N+2,1-2\ell;2\right)\;,\nonumber\\
\mathcal{A}_3^\prime&=F^{(0,1,0,0)}\left(\frac{3}{2}+\ell,2N+2\ell+4,3+2\ell;2\right)\;,\nonumber\\
\mathcal{A}_4^\prime&=2\ell F\left(-\frac{1}{2}-\ell,2N+2,-2\ell;2\right)\nonumber\\&-2(N+1)F\left(\frac{1}{2}-\ell,2N+3,1-2\ell;2\right)\;.\nonumber
\end{align}
Here,  $F^{(0,1,0,0)}(a,b,c;z)$ represents the first derivative of the hypergeometric function with respect to the second argument.

Observe from Eq.~\eqref{generaldelta} that $\bar{\omega}$ appears in this equation; but $\mbox{Re}\delta_j$ does \textit{not} change the sign when $\bar{\omega}<0$. This fact becomes more clear by expanding Eq.~\eqref{generaldelta} around small $\bar{\omega}$. The leading order gives
\begin{equation}
\mbox{Re}\delta_j=-\dfrac{\sigma_j}{4\pi}\left(\dfrac{\ell!}{(2\ell)!}\right)^2\prod^{\ell-1/2}_{p=1}(p^2+4\bar{\omega}^2)\left(\dfrac{r_+-r_-}{L}\right)^{2\ell+1}\;.\label{finaldelta}
\end{equation}
Now it becomes obvious the imaginary part of $\omega_j$ does not become positive. Hence, the Dirac field \textit{does not} trigger superradiant instabilities in the regime $\omega r_+<qQ$.

These analytic calculations are applicable only for small AdS BHs and in the low frequency limit. In order to study charged Dirac QNMs in a larger parameter space, one has to employ numerical methods, which will be addressed in the next section.

\section{Numeric calculations}
\label{secnum}
In this section, we present numeric calculations for the charged Dirac quasinormal frequencies on RN-AdS BHs, with the \textit{vanishing energy flux} boundary conditions. To achieve this goal, we first briefly introduce the numeric methods, and then apply them to several specific examples to display the effect of the electric charge on Dirac QNMs.

\subsection{Numeric method}
In this subsection, we illustrate three different numeric methods that we have employed to solve the eigenfrequency $\omega$: a direct integration method, the Horowitz-Hubeny method, and a pseudospectral method. The first two methods have already been utilized to study the neutral Dirac fields~\cite{Wang:2017fie}, we therefore put more details on the third approach.

\subsubsection{Direct integration method}
In order to obtain quasinormal frequencies for charged Dirac fields, a direct integration method may be utilized. For numerical convenience, we first rewrite the second order radial equation~\eqref{ChargedDiracU2radialR1} in a first order form, following~\cite{Herdeiro:2011uu,Wang:2012tk,Wang:2014eha,Wang:2015fgp}. For that purpose, we shall define two new fields $\left\{\chi,\psi\right\}$, which will asymptote to $\left\{\alpha_1,\beta_1\right\}$, at infinity. Then from Eq.~\eqref{ChargedasysolR1}, these new fields are related with the original field and its derivative through the transformation
\begin{equation}
\left(\begin{array}{c}R_1 \vspace{2mm}\\ \dfrac{dR_1}{dr}\end{array}\right)= \left(\begin{array}{cc} 1 & \frac{L}{r} \vspace{2mm}\\ 0 & -\frac{L}{r^2} \end{array}\right) \left(\begin{array}{c}\chi\vspace{2mm}\\ \psi\end{array}\right) \equiv \mathbf{T} \left(\begin{array}{c}\chi\vspace{2mm}\\ \psi\end{array}\right) \;.\label{Chargedtransformation}
\end{equation}
By defining the vector $\mathbf{\Psi}^T=(\chi,\psi)$ for the new fields, and another vector $\mathbf{V}^T=(R_1,\frac{d}{dr}R_1)$ for the original field and its derivative, Eq.~\eqref{ChargedDiracU2radialR1} may be rewritten as a first order differential equation in a matrix form:
\begin{equation}
\dfrac{d\mathbf{\Psi}}{dr}=\mathbf{T}^{-1}\left(\mathbf{X}\mathbf{T}-\dfrac{d\mathbf{T}}{dr}\right) \mathbf{\Psi} \;,\label{Chargedradialmatrix}
\end{equation}
where a matrix $\mathbf{X}$ is defined through
\begin{equation}
\dfrac{d\mathbf{V}}{dr}=\mathbf{X}\mathbf{V} \; ,
\end{equation}
which can be read out from the original radial equation~\eqref{ChargedDiracU2radialR1} straightforwardly.

To solve Eq.~\eqref{Chargedradialmatrix}, we shall first initialize $R_1$ near the event horizon $r_+$, by using the Frobenius' method
\begin{equation}
R_{1}=(r-r_+)^\rho \sum_{j=0}^\infty c_j\;(r-r_+)^j\;,\label{chargedexpansionD}
\end{equation}
with
\begin{equation}
\rho=\dfrac{1}{2}-i\dfrac{(\omega+qC) r_+-qQ}{4\pi r_+T_H}\;,\label{chargedrhoD}
\end{equation}
where $T_H$ is the Hawking temperature given by Eq.~\eqref{hawkingT}. Note that the ingoing wave boundary condition has been imposed, and the expansion coefficients $c_j$ can be directly obtained by substituting Eq.~\eqref{chargedexpansionD} into Eq.~\eqref{ChargedDiracU2radialR1}. Then according to the transformation given by Eq.~\eqref{Chargedtransformation}, one may easily get the initial conditions for $\mathbf{\Psi}$. By integrating $\mathbf{\Psi}$ outwards from the event horizon through Eq.~\eqref{Chargedradialmatrix} and evaluating it at infinity, QNMs may be obtained after imposing boundary conditions at infinity, given by Eqs.~\eqref{Chargedbc1} and~\eqref{Chargedbc2}.

\subsubsection{Horowitz-Hubeny method}
The Horowitz-Hubeny method, first employed in~\cite{Horowitz:1999jd}, is an efficient approach to calculate QNM frequencies for asymptotically AdS BHs. The standard procedures for this method are as follows. First, we shall transform Eq.~\eqref{ChargedDiracU2radialR1} into the Schr\"odinger-like form
\begin{equation}
\dfrac{d^2\phi_1}{dr_\ast^2}+\Big((\omega+qC)^2-V\Big)\phi_1=0\;,\label{ChargedSchroeq}
\end{equation}
with
\begin{align}
&V=iqQ\dfrac{\Delta_r}{r^4}-\dfrac{ir^2}{2}\left(\omega+qC-\dfrac{qQ}{r}\right)\left(\dfrac{\Delta_r}{r^4}\right)^\prime+\dfrac{\lambda^2}{r^4}\Delta_r\nonumber\\&+\dfrac{2(\omega+qC) qQ}{r}-\dfrac{q^2Q^2}{r^2}-\dfrac{2\Delta_r^2}{r^6}
-\dfrac{\Delta_r}{4}\left(\dfrac{\Delta_r^\prime}{r^4}\right)^\prime+\dfrac{\Delta_r^{\prime2}}{16r^4}
\;,\nonumber
\end{align}
where $\phi_1$ is related to $R_1$ by
\begin{equation}
\phi_1=\dfrac{r}{\Delta_r^{1/4}}R_1\; ,\label{chargedtransf0}
\end{equation}
and the tortoise coordinate $r_\ast$ is defined as
\begin{equation}
\dfrac{dr_\ast}{dr}=\dfrac{r^2}{\Delta_r} \ .\label{tortoisecoor}
\end{equation}

By analyzing the asymptotic behavior of $\phi_1$ close to the event horizon $r_+$, from Eq.~\eqref{ChargedSchroeq} we get
\begin{equation}
\phi_1\thicksim e^{\pm i\varpi r_\ast}\;,\nonumber
\end{equation}
with
\begin{equation}
\varpi=\omega+qC-\dfrac{qQ}{r_+}+\dfrac{i}{4r_+}\left(1+\dfrac{3r_+^2}{L^2}-\dfrac{Q^2}{r_+^2}\right)\;,\label{chargedvarpi}
\end{equation}
where $-/+$ sign stands for ingoing/outgoing waves.
Then we further make the following transformation for the radial function
\begin{equation}
\phi_1=e^{-i\varpi r_\ast}\Phi_1\;,\label{chargedtransf}
\end{equation}
and change the variable from $r$ to $x$ through $x=1/r$; Eq.\eqref{ChargedSchroeq} becomes
\begin{equation}
S(x)\dfrac{d^2\Phi_1}{dx^2}+\dfrac{T(x)}{x-x_+}\dfrac{d\Phi_1}{dx}+\dfrac{U(x)}{(x-x_+)^2}\Phi_1=0\;,\label{ChargedHHeq}
\end{equation}
where the polynomials $S, T, U$ are defined in the Appendix~\ref{app1}. Two comments for the above equation are in order: (1) for the radial function $\Phi_1$, the ingoing wave boundary condition at the event horizon is automatically satisfied; (2) in terms of $x$, the entire space outside the event horizon $r_+<r<\infty$ is mapped into a finite region $0<x<x_+$, with $x_+=1/r_+$.

To evaluate QNMs by using Horowitz-Hubeny method, we shall expand all functions in Eq.~\eqref{ChargedHHeq} around $x_+$,
\begin{align}
&\Phi_1=\sum_{j=0}^\infty a_j(x-x_+)^j\;,\;\;\;S(x)=\sum_{n=0}^6 s_n(x-x_+)^n\;,\nonumber\\
&T(x)=\sum_{n=0}^6 t_n(x-x_+)^n\;,U(x)=\sum_{n=0}^6 u_n(x-x_+)^n\;,\nonumber%
\end{align}
where the recurrence relations for $a_j$ are given in Appendix~\ref{app1}, and the expansion coefficients $\{s_n, t_n, u_n\}$ can be read off from Eq.~\eqref{ChargedSeriesfunc} straightforwardly.

The boundary conditions derived in Eqs.~\eqref{Chargedbc1} and \eqref{Chargedbc2} for $R_1$, by using Eqs.~\eqref{chargedtransf0} and~\eqref{chargedtransf}, become
\begin{equation}
\sum_ja_j(-x_+)^j\left(1+\dfrac{j}{\gamma x_+}\right)=0\;,\label{BCHH}
\end{equation}
for $\Phi_1$, with
\begin{equation}
\gamma=\gamma_1\equiv iL\left(\ell+\frac{1}{2}-(\omega+qC+\varpi)L\right)\;,\label{HHbc1}
\end{equation}
for the first boundary condition, and
\begin{equation}
\gamma=\gamma_2\equiv-iL\left(\ell+\frac{1}{2}+(\omega+qC+\varpi)L\right)\;,\label{HHbc2}
\end{equation}
for the second boundary condition, where $\varpi$ is given by Eq.~\eqref{chargedvarpi}.

\subsubsection{Pseudospectral method}
To solve boundary values problems, a pseudospectral method may alternatively be applied, see $e.g.$~\cite{trefethen2000spectral}. The second order differential equation~\eqref{ChargedDiracU2radialR1} is a quadratic eigenvalue problem. For numeric convenience, we first transform this equation into a linear eigenvalue problem, through
\begin{equation}
R_1=\dfrac{\Delta_r^{1/4}}{r}e^{-i\varpi r_\ast}\tilde{R}_1\;,\label{spectraltrans}
\end{equation}
where the tortoise coordinate $r_\ast$ is defined in Eq.~\eqref{tortoisecoor}, and $\varpi$ is given in Eq.~\eqref{chargedvarpi}. By letting
\begin{equation}
z=1-\dfrac{2r_+}{r}\;,\label{rtoz}
\end{equation}
the integration domain changes from $r\in[r_+,\infty]$ to $z\in[-1,+1]$. With the above two transformations in mind, Eq.~\eqref{ChargedDiracU2radialR1} turns into the following form
\begin{equation}
\mathcal{B}_0(z,\omega)\tilde{R}_1(z)+\mathcal{B}_1(z,\omega)\tilde{R}^\prime_1(z)+\mathcal{B}_2(z,\omega)\tilde{R}^{\prime\prime}_1(z)=0\; .\label{spectraleq1}
\end{equation}
Here $\prime$ denotes derivative with respect to $z$. Each of the $\mathcal{B}_j (j=1,2,3)$ can be derived straightforwardly (by substituting Eqs.~\eqref{spectraltrans} and~\eqref{rtoz} into Eq.~\eqref{ChargedDiracU2radialR1}) and they are linear in $\omega$, i.e. $\mathcal{B}_j(z,\omega)=\mathcal{B}_{j,0}(z)+\omega\mathcal{B}_{j,1}(z)$.

The pseudospectral method solves a differential equation by replacing a continuous variable with a set of discrete grid points. For this purpose, we introduce the Chebyshev points
\begin{equation}
z_j=\cos\left(\dfrac{j\pi}{n}\right)\;,\;\;\;\;\;\;j=0,1,...,n\;,\label{spectralpoints}
\end{equation}
where $n$ denotes the number of grid points. One may use these points to construct Chebyshev differentiation matrices~\cite{trefethen2000spectral} and apply these matrices to differentiate $\tilde{R}_1(z)$. Then the differential equation~\eqref{spectraleq1} becomes an algebraic equation
\begin{equation}
(M_0+\omega M_1)\tilde{R}_1(z)=0\;,\label{spectraleq2}
\end{equation}
where $M_0$ and $M_1$ are matrices, $(M_0)_{ij}=\mathcal{B}_{0,0}(z_i)\delta_{ij}+\mathcal{B}_{1,0}(z_i)D^{(1)}_{ij}+\mathcal{B}_{2,0}(z_i)D^{(2)}_{ij}$, and similarly for $M_1$. For simplicity, we define the second order Chebyshev differential matrix $D^{(2)}$ by squaring the first order Chebyshev differential matrix $D^{(1)}$~\cite{trefethen2000spectral}.

To solve the eigenvalue equation~\eqref{spectraleq2}, one has to impose proper boundary conditions for $\tilde{R}_1$. At the horizon, from Eq.~\eqref{spectraltrans} we impose regular boundary condition, since ingoing wave boundary condition is satisfied automatically for $R_1$. At infinity, from Eq.~\eqref{spectraltrans} one may obtain
\begin{equation}
\dfrac{\tilde{R}^\prime_1}{\tilde{R}_1}=\dfrac{i\varpi L^2}{2r_+}-\dfrac{L}{2r_+}\left(\dfrac{\alpha_1}{\beta_1}\right)^{-1}\;,\label{spectralbc}
\end{equation}
where $\alpha_1/\beta_1$ is given in Eqs.~\eqref{Chargedbc1} and~\eqref{Chargedbc2}, corresponding to the two sets of boundary conditions.

\subsection{Numeric results}

With the above numeric methods at hand, in this part we present numeric results for charged Dirac QNM frequencies on RN-AdS BHs, beyond the small BH and low frequency approximations.

In the numeric calculations, all physical quantities are normalised by the AdS radius $L$, which amounts to setting $L=1$, without loss of generality. Since the gauge constant $C$ in the electrostatic potential, given by Eq.~\eqref{potential}, only shifts the real part of QNMs by $\Re(\omega)\rightarrow\Re(\omega)-qC$, we simply choose this constant as $C=0$ in the numeric calculations. Moreover, we use $\omega_1$ ($\omega_2$) to represent the quasinormal frequency corresponding to the first (second) boundary condition. Most of the numeric results presented in this part are generated by a pseudospectral method, and they are also double checked by a direct integration and the Horowitz-Hubeny methods when they are applicable.

\begin{figure*}
\begin{center}
\begin{tabular}{c}
\includegraphics[clip=false,width=0.396\textwidth]{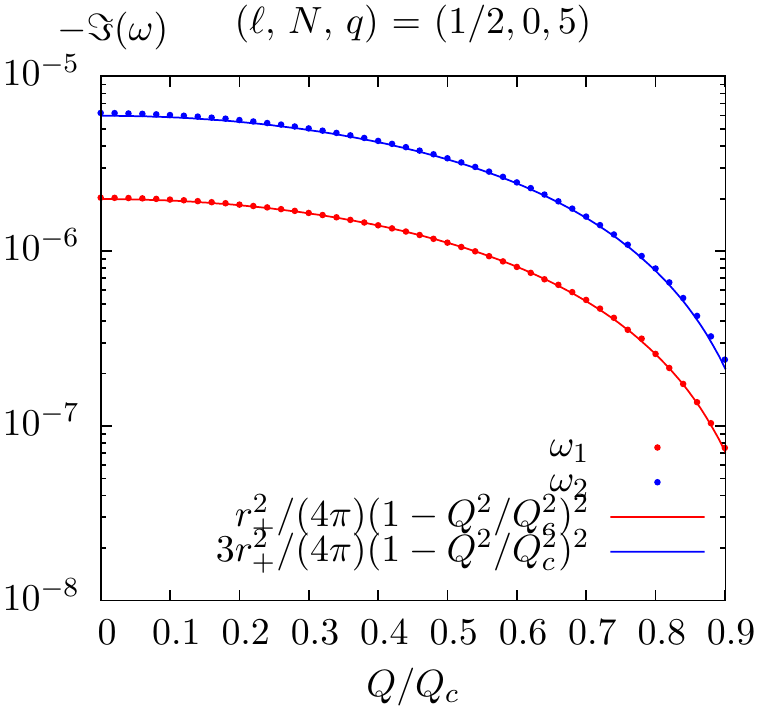}\;\;\;\;\;\;\;\;\;\;\;\;\;\;\;\;\;\;\;\;\;\;\;\includegraphics[clip=false,width=0.396\textwidth]{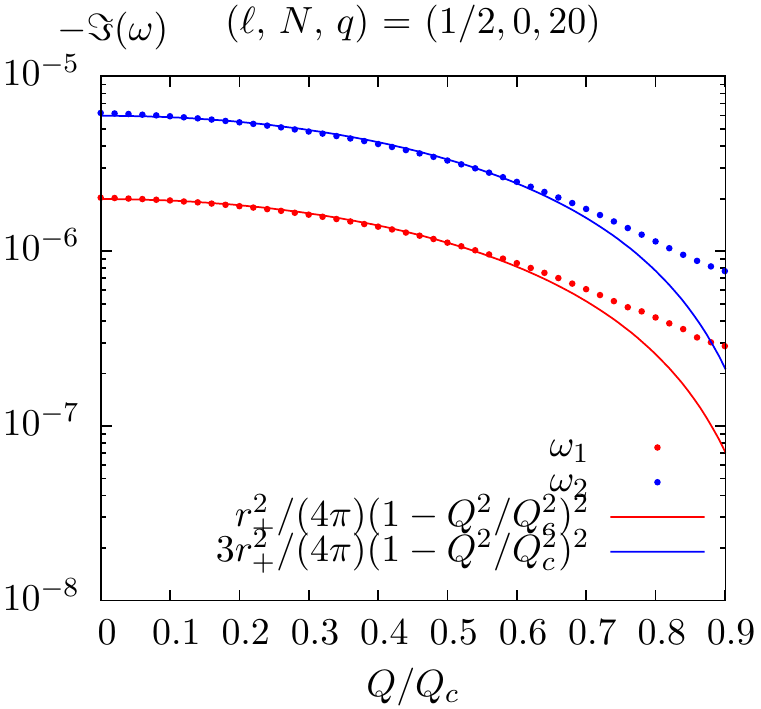}
\end{tabular}
\end{center}
\caption{\label{AVNF} (color online). Analytic (thin solid lines) and numeric (dots)  imaginary part of QNM frequencies, $vs.$ $Q/Q_c$ for the fundamental modes ($N=0$) with the first (red) and second (blue) boundary conditions, for a small BH with $r_+=0.005$ with different field charges $q=5$ (left) and $q=20$ (right). Observe this figure is made with semilogarithmic coordinates.}
\end{figure*}

We start with a comparison between analytic and numeric calculations for small BHs in Fig.~\ref{AVNF}. This comparison may be used not only to verify the validity of the analytic calculations but also as another check on the numerics. We have fixed the overtone number $N=0$ and the angular momentum quantum number $\ell=1/2$. Under these conditions and under the small BH approximation, the imaginary part of QNMs given in Eq.~\eqref{finaldelta} becomes
\begin{equation}
\mbox{Re}\delta_1=-\dfrac{r_+^2}{4\pi}\left(1-\dfrac{Q^2}{Q_c^2}\right)^2\;,\;\;\mbox{Re}\delta_2=-\dfrac{3r_+^2}{4\pi}\left(1-\dfrac{Q^2}{Q_c^2}\right)^2\;,\label{deltalimit}
\end{equation}
up to leading order of $r_+$, and where the background charge $Q$ is constrained by the extremal charge $Q_c$ as given in Eq.~\eqref{criticalQ}. The field charge $q$ does not show up in the above formulas, and when $Q=0$ neutral Dirac results are recovered~\cite{Wang:2017fie}. Furthermore, from the above equation, it seems that Re$\delta_j (j=1, 2)$ becomes \textit{zero} (i.e. imaginary part of QNMs becomes \textit{zero}) when a small BH approaches extremality, implying the existence of arbitrarily long-lived modes. This is reminiscent of the marginal scalar and Proca clouds discussed in~\cite{Degollado:2013eqa,Sampaio:2014swa}, arbitrarily long lived bosonic modes near extremal BHs for bosonic fields. In order to verify the above analytic formulas, we have generated numeric data for a small BH with size $r_+=0.005$. As one may observe from Fig.~\ref{AVNF}, the analytic formulas given in Eq.~\eqref{deltalimit} match well with numeric data for both the field charge $q=5$ (when $Q$ varies from $0$ to $0.9Q_c$) and the field charge $q=20$ (when $Q$ varies from $0$ to $0.6Q_c$). By comparing these two cases, one may conclude that the smaller the charge coupling $qQ$, the better agreement between analytic and numeric results.

Beyond the small BH approximation, one has to resort to numeric methods to calculate QNMs. By employing the numeric methods we introduced in the above subsection, a few selection numeric data of charged Dirac QNMs for different BH size $r_+$, angular momentum quantum number $\ell$ and overtone number $N$ are presented below.

\begin{figure*}
\begin{center}
\begin{tabular}{c}
\includegraphics[clip=false,width=0.396\textwidth]{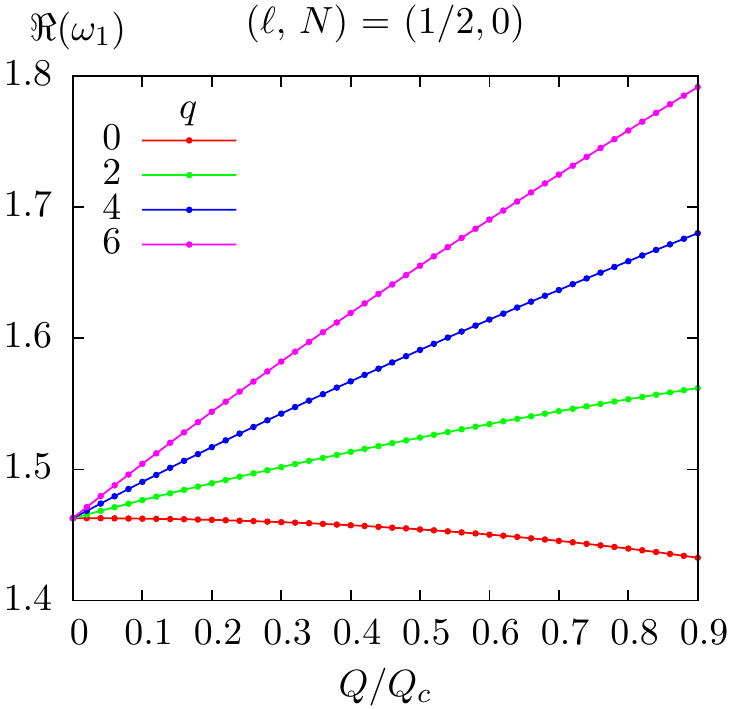}\;\;\;\;\;\;\;\;\;\;\;\;\;\;\;\;\;\;\;\;\;\;\;\includegraphics[clip=false,width=0.396\textwidth]{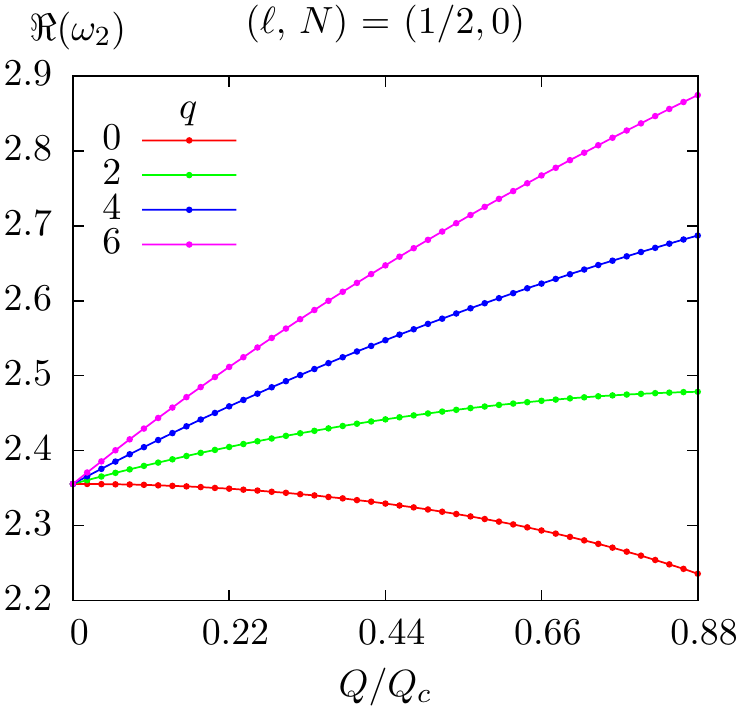}
\\
\includegraphics[clip=false,width=0.396\textwidth]{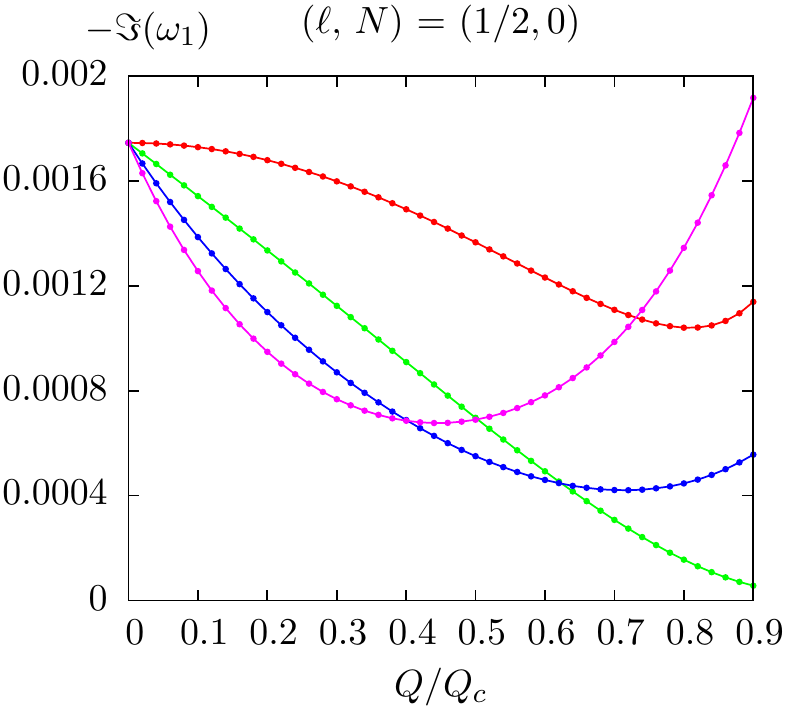}\;\;\;\;\;\;\;\;\;\;\;\;\;\;\;\;\;\;\;\;\;\;\;\includegraphics[clip=false,width=0.396\textwidth]{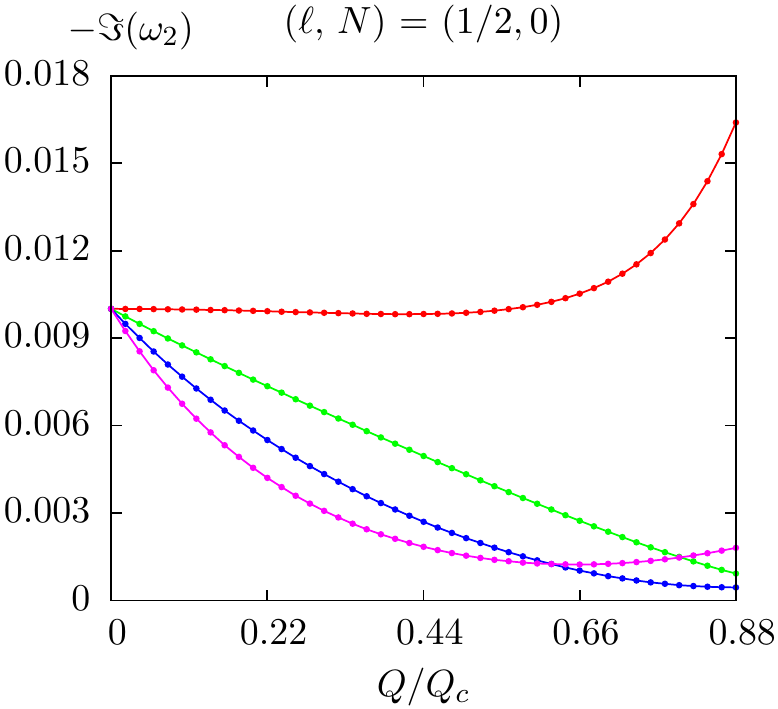}
\end{tabular}
\end{center}
\caption{\label{chargeQNMsrp01} (color online). Real (top) and imaginary (bottom) parts of QNMs for charged Dirac fields $vs.$ $Q/Q_c$ for BH size $r_+=0.1$, $N=0$ and $\ell=1/2$, with the first (left) and second (right) boundary conditions. We have chosen the field charge values $q=0,2,4,6$. The figure keys in the top panels also apply to the bottom panels.}
\end{figure*}

\begin{figure*}
\begin{center}
\begin{tabular}{c}
\includegraphics[clip=false,width=0.396\textwidth]{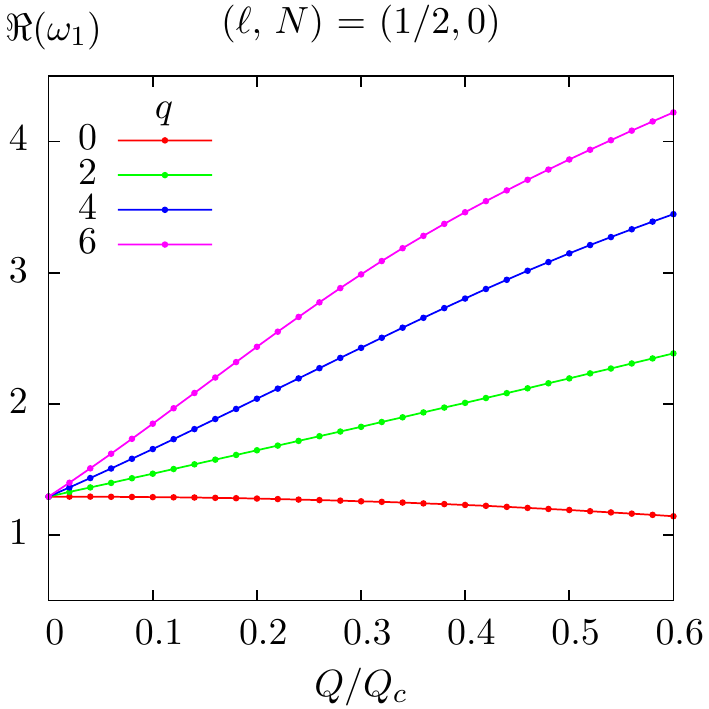}\;\;\;\;\;\;\;\;\;\;\;\;\;\;\;\;\;\;\;\;\;\;\;\includegraphics[clip=false,width=0.396\textwidth]{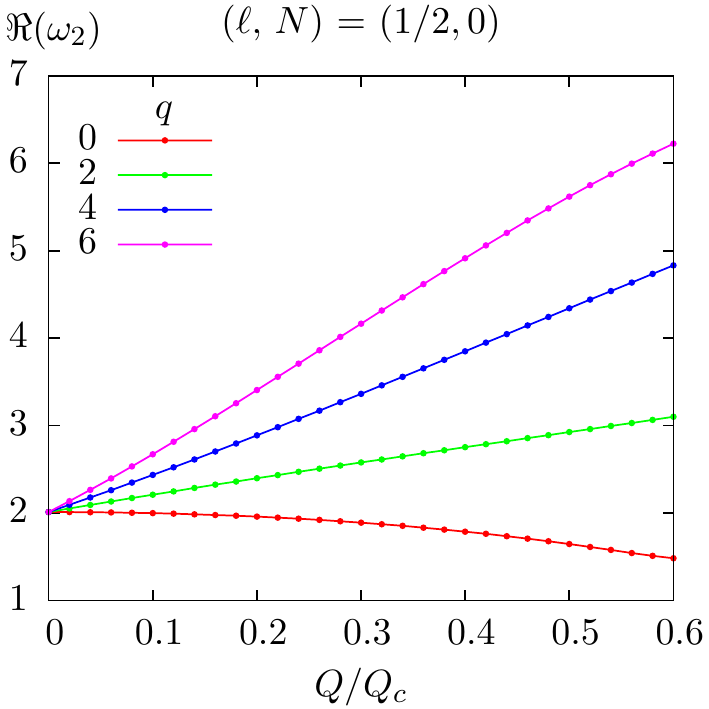}
\\
\includegraphics[clip=false,width=0.396\textwidth]{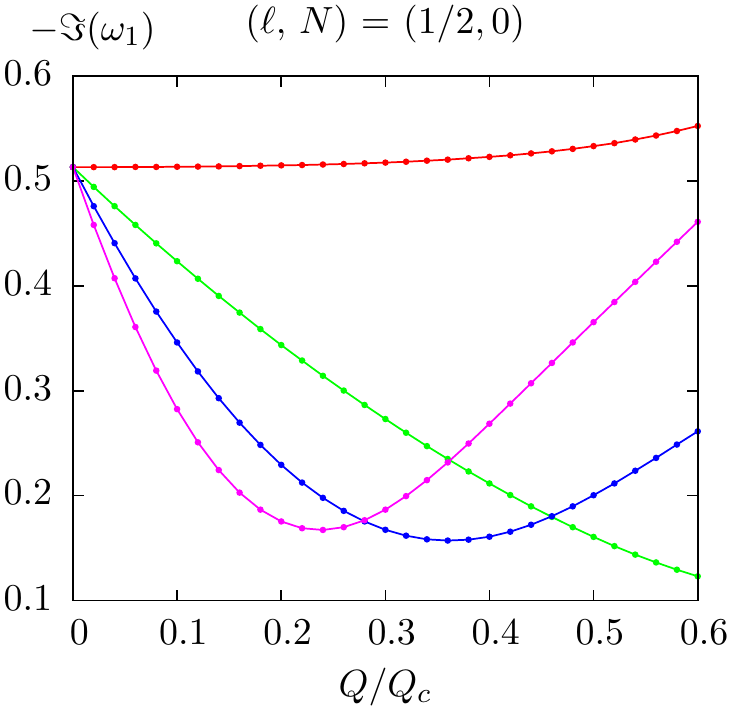}\;\;\;\;\;\;\;\;\;\;\;\;\;\;\;\;\;\;\;\;\;\;\;\includegraphics[clip=false,width=0.396\textwidth]{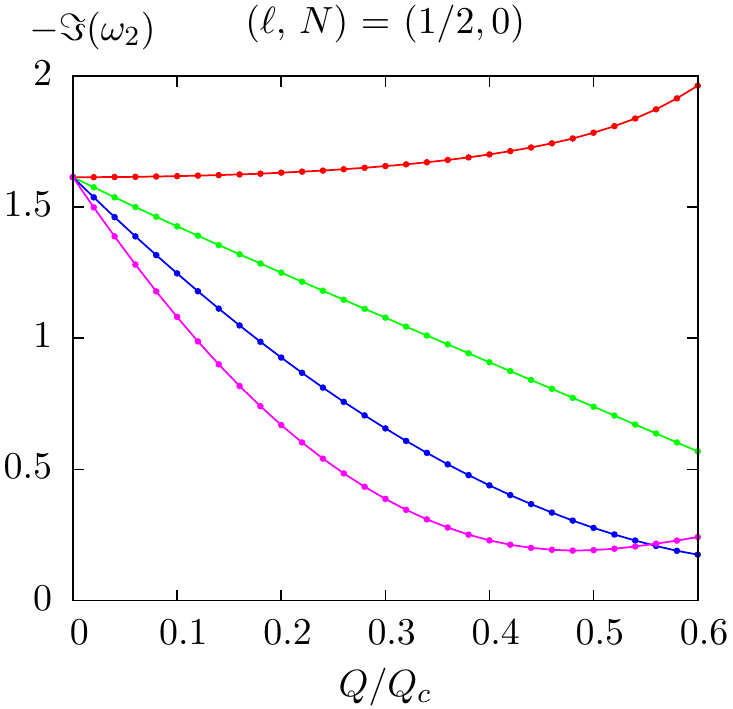}
\end{tabular}
\end{center}
\caption{\label{chargeQNMsrp1} (color online). Real (top) and imaginary (bottom) parts of QNMs for charged Dirac fields are presented in terms of $Q/Q_c$ for BH size $r_+=1$, the overtone number $N=0$ and the angular momentum quantum number $\ell=1/2$, with the first (left) and second (right) boundary conditions. We have chosen field charge as $q=0$, $q=2$, $q=4$, and $q=6$. The legends in the top panels apply to the bottom panels as well.}
\end{figure*}

\begin{figure*}
\begin{center}
\begin{tabular}{c}
\includegraphics[clip=false,width=0.396\textwidth]{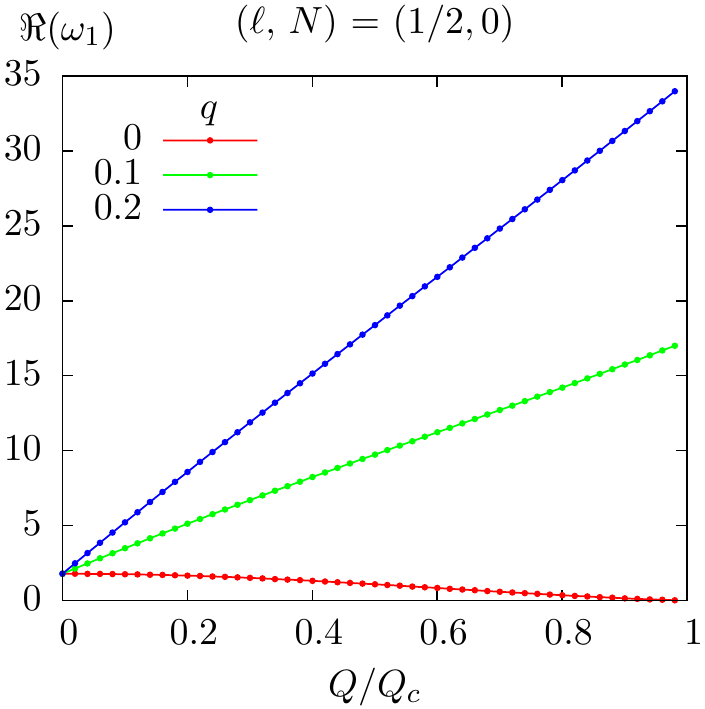}\;\;\;\;\;\;\;\;\;\;\;\;\;\;\;\;\;\;\;\;\;\;\;\includegraphics[clip=false,width=0.396\textwidth]{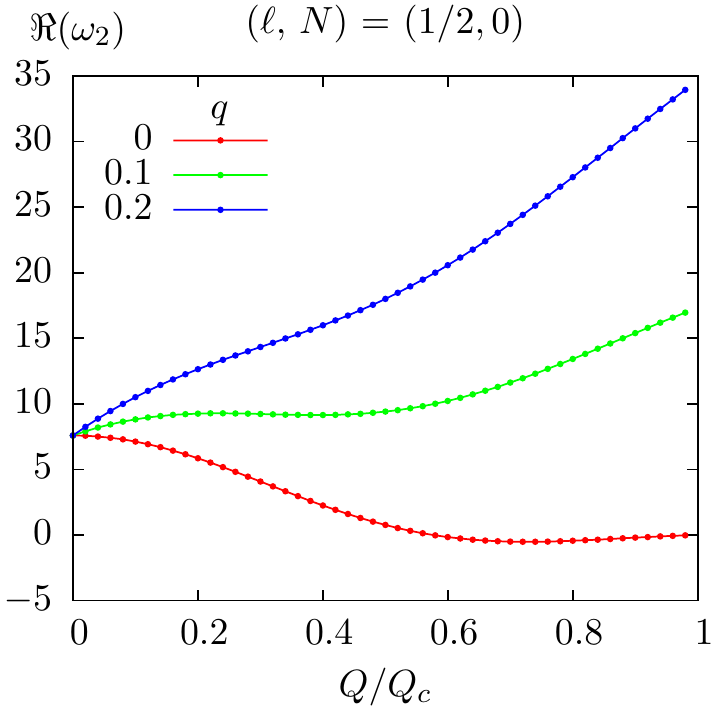}
\\
\includegraphics[clip=false,width=0.396\textwidth]{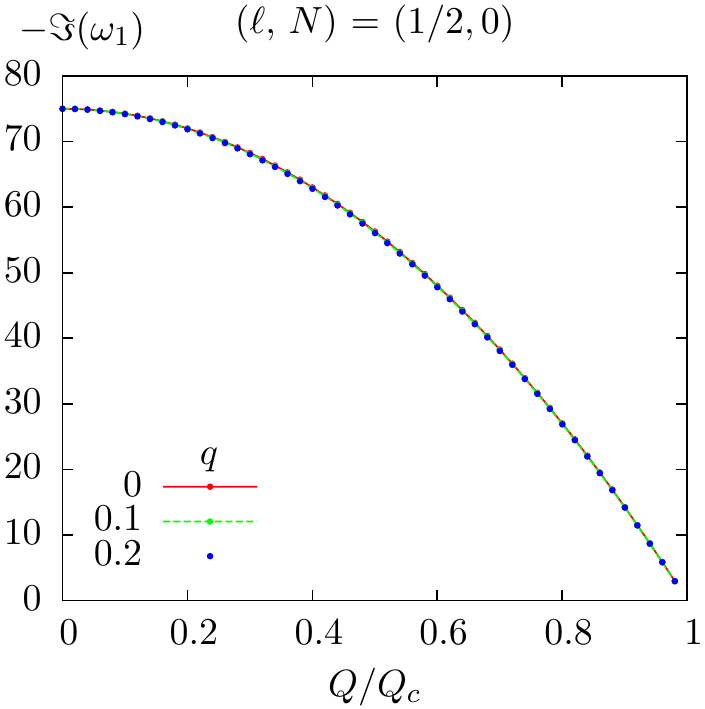}\;\;\;\;\;\;\;\;\;\;\;\;\;\;\;\;\;\;\;\;\;\;\;\includegraphics[clip=false,width=0.396\textwidth]{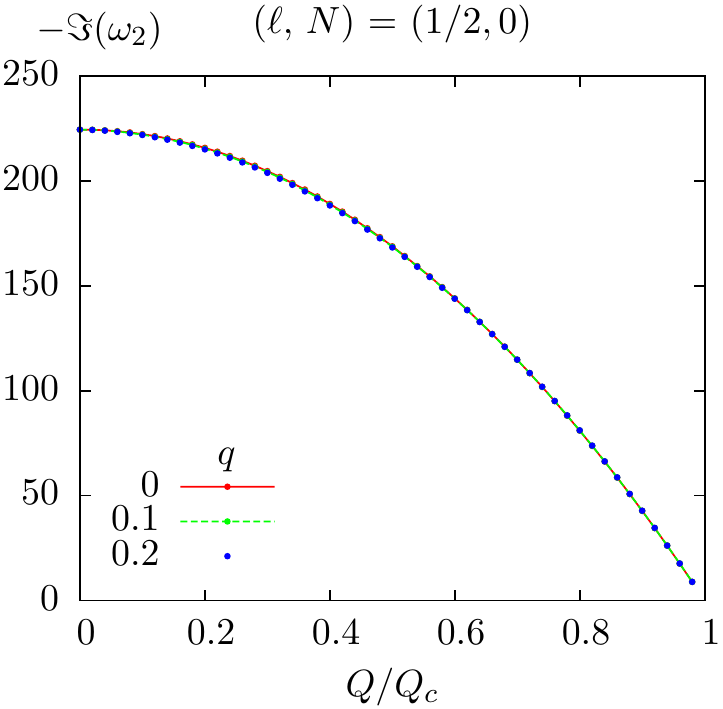}
\end{tabular}
\end{center}
\caption{\label{chargeQNMsrp100} (color online). Real (top) and imaginary (bottom) parts of QNMs for charged Dirac fields $vs.$ $Q/Q_c$ for BH size $r_+=100$, $N=0$ and $\ell=1/2$, with the first (left) and second (right) boundary conditions. We have chosen field charge as $q=0,0.1,0.2$.}
\end{figure*}

We first study charge effects (both for the background and the field charges) on QNMs for different BH sizes, by fixing the angular momentum quantum number $\ell (=1/2)$ and the overtone number $N (=0)$, with two Robin boundary conditions. We have chosen three different BH sizes as $r_+=0.1/1/100$, and the corresponding numeric results are shown in Figs.~\ref{chargeQNMsrp01}$-$\ref{chargeQNMsrp100}. Note that in all of these figures, real (top panels) and imaginary (bottom panels) part of QNMs are presented in terms of $Q/Q_c$, with the first (left panels) and second (right panels) boundary conditions.

The effect of the background electric charge on the QNMs, for all three BH sizes, may be appreciated by observing the $q=0$ curves, in Figs.~\ref{chargeQNMsrp01}$-$\ref{chargeQNMsrp100}. For the case $r_+=0.1$, exhibited in Fig.~\ref{chargeQNMsrp01}, the real part of the quasinormal frequency, for both boundary conditions, decreases as the background charge $Q$ increases; the magnitude of the imaginary part, on the other hand, initially decreases and subsequently increases, as $Q$ increases, for both boundary conditions\footnote{This is less obvious but true, by observing the data, for the second boundary condition.}. For the case $r_+=1$, exhibited in Fig.~\ref{chargeQNMsrp1}, the real part of QNMs behaves similarly to the previous case; the magnitude of the imaginary part, on the other hand, now increase as $Q$ increases for both boundary conditions. For the case $r_+=100$, shown in Fig.~\ref{chargeQNMsrp100}, the real part of QNMs decreases as $Q$ increases for the first boundary condition only; for the second it first decreases and then increases. The magnitude of the imaginary part decreases as $Q$ increases for both boundary conditions.

Turning on the field charge $q$, the real part of QNMs increases with increasing $Q$, for both boundary conditions and all three BH sizes, except for the case of $r_+=100$ and $q=0.1$ with the second boundary condition. For this case, the real part of QNMs first increases then decreases and finally increases with increasing $Q$. For larger $q$ (say $q=0.2$), the aforementioned behavior, i.e. the real part of QNMs increases with increasing $Q$, is recovered. The magnitude of the imaginary part, on the other hand, for both boundary conditions and for the cases $r_+=0.1$ and $r_+=1$, decreases with increasing $Q$ when $q=2$. When $q$ becomes larger, one observes that the magnitude of the imaginary part  first decreases and then increases as $Q$ increases, for both boundary conditions. For the case of $r_+=100$, the magnitude of the imaginary part  with both boundary conditions decreases as $Q$ increases.

\begin{figure*}
\begin{center}
\begin{tabular}{c}
\includegraphics[clip=false,width=0.381\textwidth]{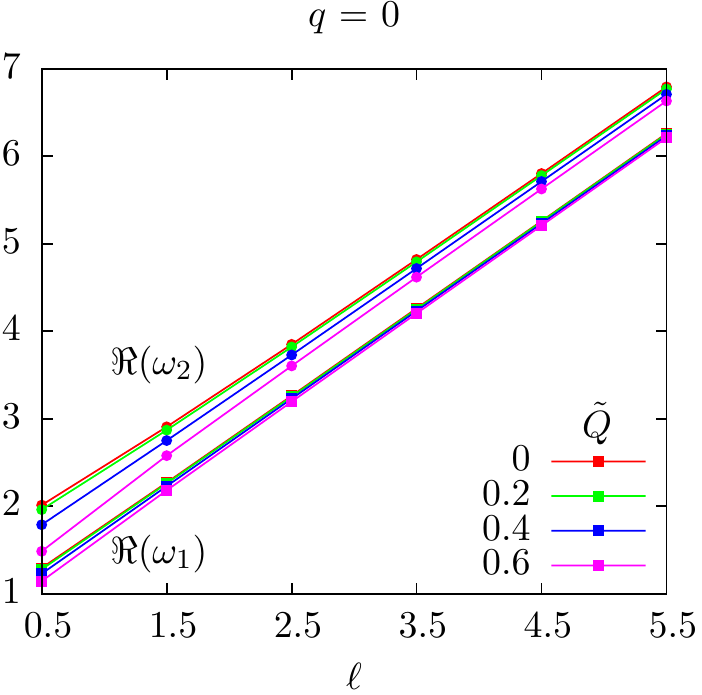}\;\;\;\;\;\;\;\;\;\;\;\;\;\;\;\;\;\;\;\;\;\;\;\includegraphics[clip=false,width=0.396\textwidth]{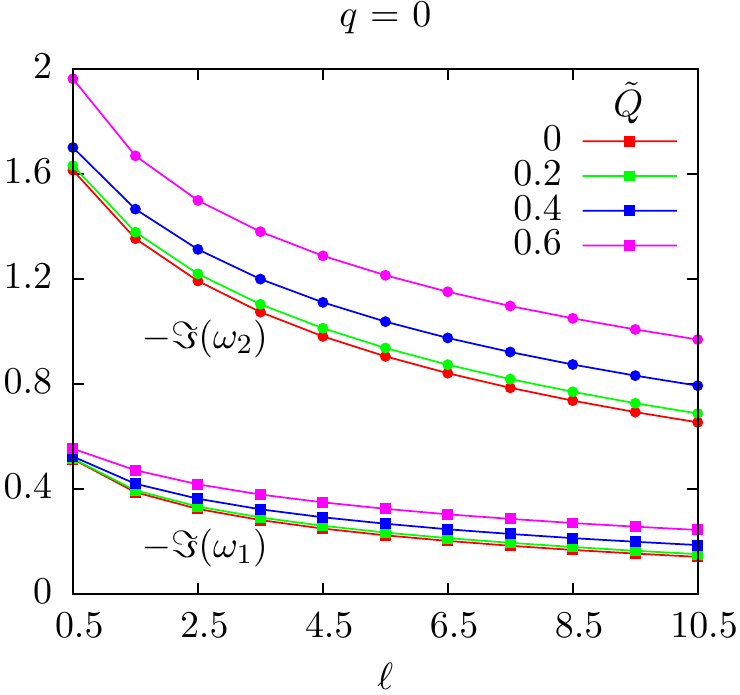}
\\
\includegraphics[clip=false,width=0.381\textwidth]{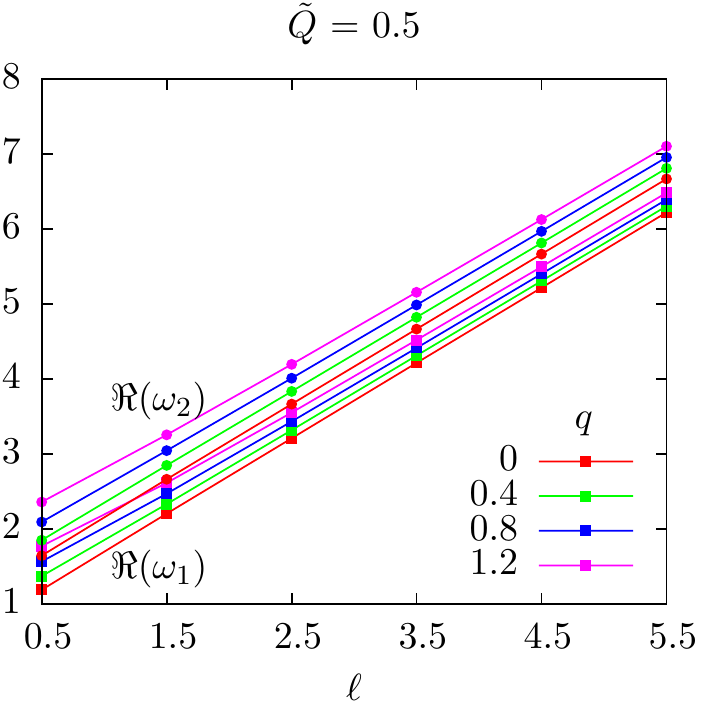}\;\;\;\;\;\;\;\;\;\;\;\;\;\;\;\;\;\;\;\;\;\;\;\includegraphics[clip=false,width=0.396\textwidth]{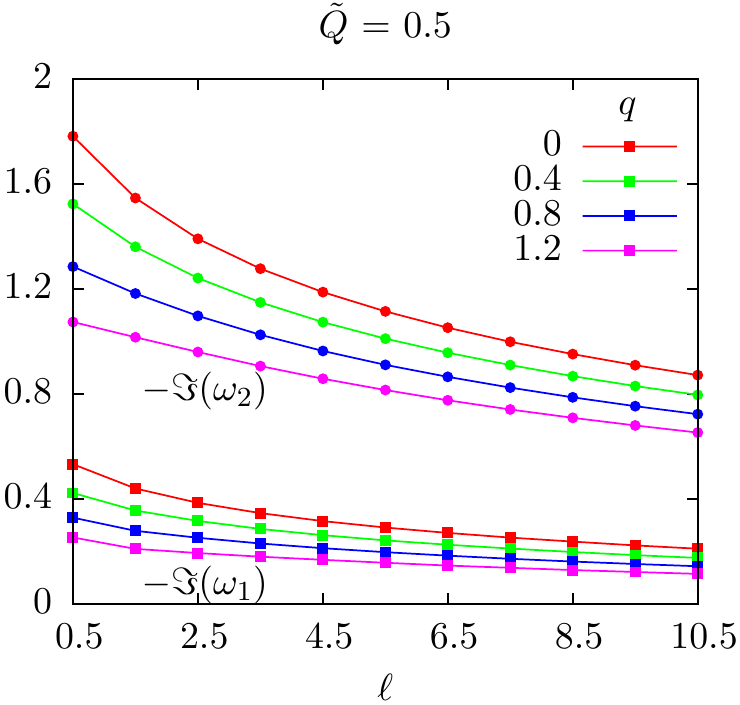}
\end{tabular}
\end{center}
\caption{\label{chargeDiracelleffects} (color online). Real (left) and imaginary (right) parts of QNMs for charged Dirac fields  $vs.$ $\ell$, by fixing the field charge $q$ (top) and the background charge $Q$ (bottom), for BH size $r_+=1$ and $N=0$. As usual, $\omega_1$ and $\omega_2$ represent frequencies with the first (solid line with square dots) and second (solid line with circle dots) boundary conditions, the same color stands for the same charge parameters $\tilde{Q}$ (top) and $q$ (bottom), and $\tilde{Q}=Q/Q_c$.}
\end{figure*}

The dependence of QNMs on $\ell$ is illustrated in Fig.~\ref{chargeDiracelleffects}, by taking an intermediate BH ($r_+=1$) with $N=0$ as an example. To show the background charge effect, we have fixed $q=0$ and varied the background charge as $Q=0$, $Q=0.2Q_c$, $Q=0.4Q_c$, $Q=0.6Q_c$ in the top panels; while to show the field charge effect, we have fixed $Q=0.5Q_c$, and varied the field charge as $q=0$, $0.4$, $0.8$, $1.2$ in the bottom panels. As one may observe, the real part of the QNMs (left panels) for both boundary conditions increase roughly linearly with increasing $\ell$; while the magnitude of the imaginary part  (right panels) decreases weakly, for both cases, with varying background charge and field charge. The panels also demonstrate how these trends are affected by varying $q$ and $Q$.

\begin{figure*}
\begin{center}
\begin{tabular}{c}
\includegraphics[clip=false,width=0.396\textwidth]{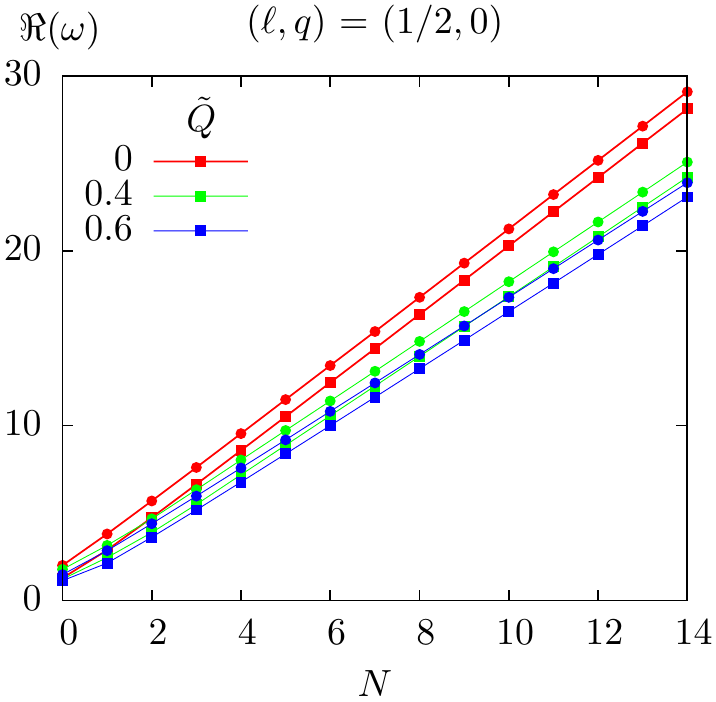}\;\;\;\;\;\;\;\;\;\;\;\;\;\;\;\;\;\;\;\;\;\;\;\includegraphics[clip=false,width=0.396\textwidth]{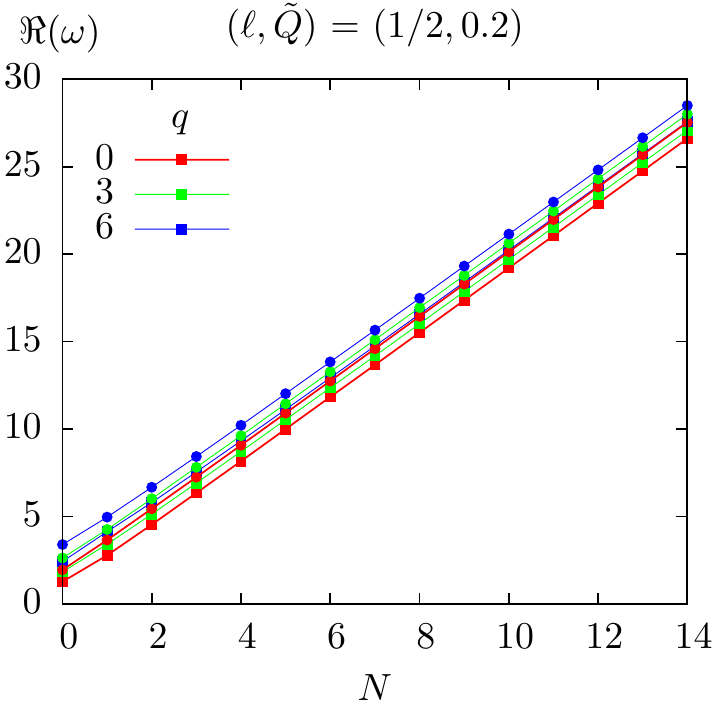}
\\
\includegraphics[clip=false,width=0.396\textwidth]{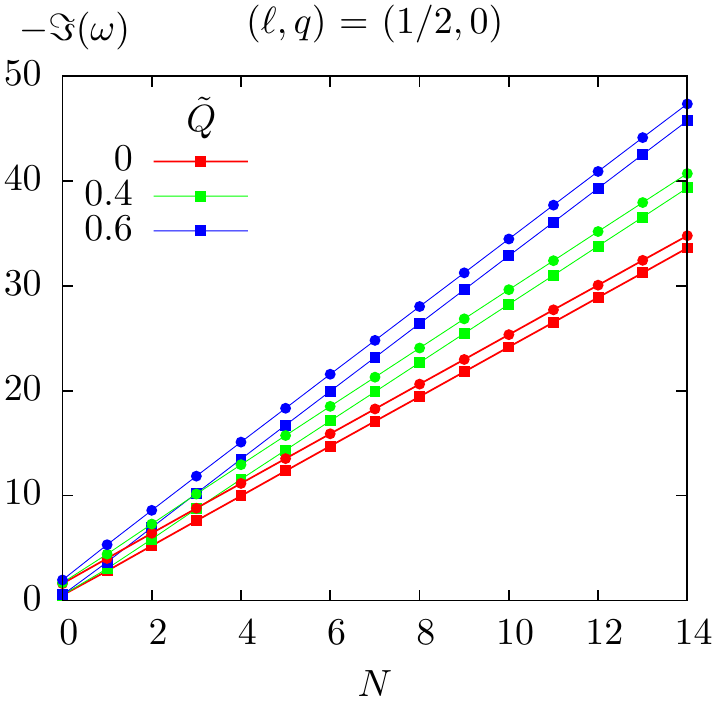}\;\;\;\;\;\;\;\;\;\;\;\;\;\;\;\;\;\;\;\;\;\;\;\includegraphics[clip=false,width=0.396\textwidth]{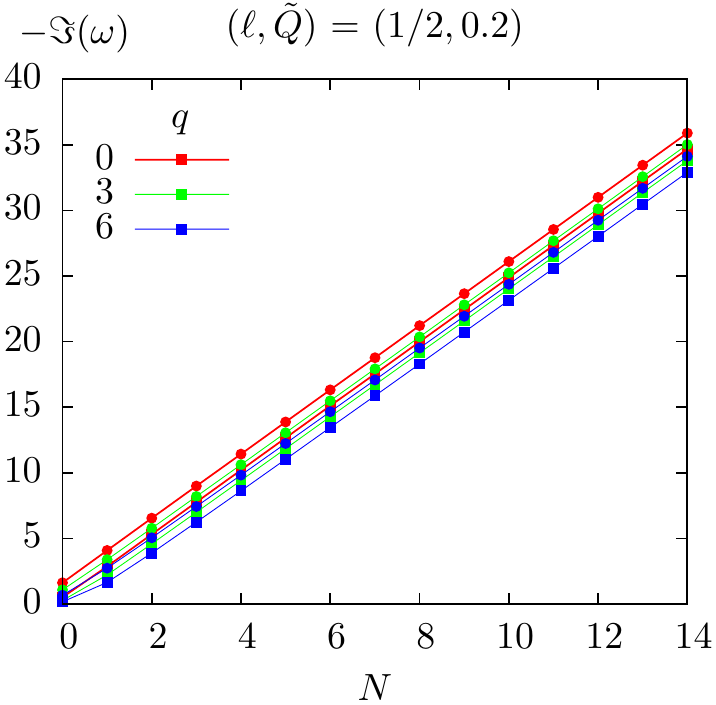}
\\
\includegraphics[clip=false,width=0.396\textwidth]{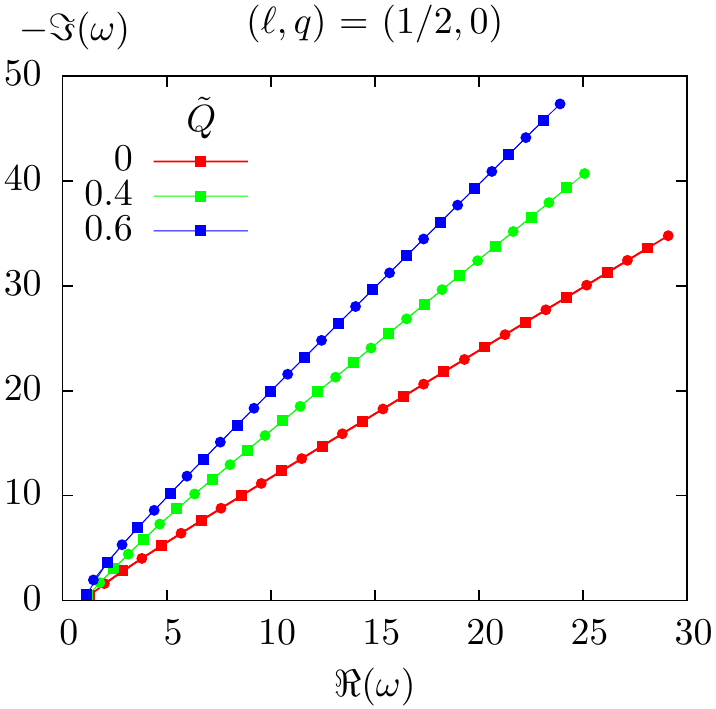}\;\;\;\;\;\;\;\;\;\;\;\;\;\;\;\;\;\;\;\;\;\;\;\includegraphics[clip=false,width=0.396\textwidth]{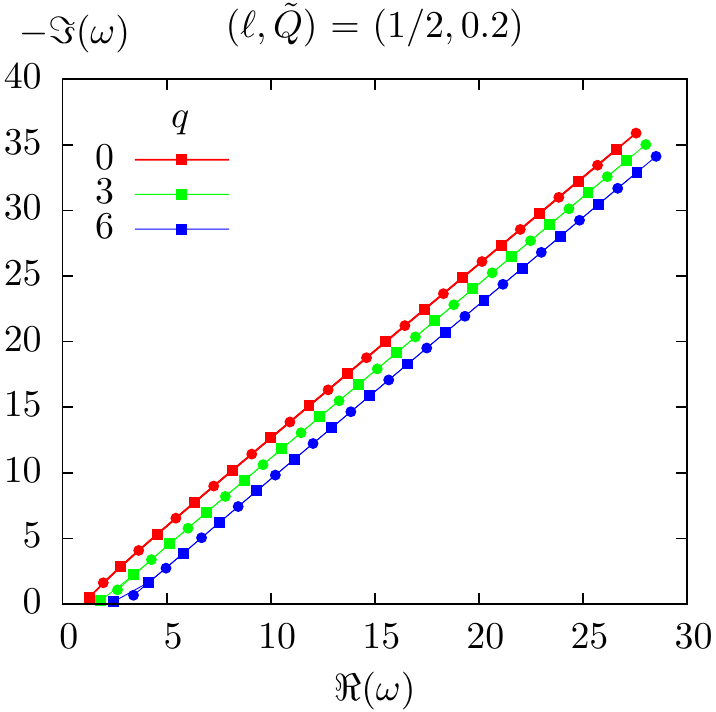}
\end{tabular}
\end{center}
\caption{\label{chargeDiracNeffects} (color online). The background charge (left panels) and field charge (right panels) effects on QNMs of charged Dirac fields, in terms of the overtone number $N$, for BH size $r_+=1$ and the angular momentum quantum number $\ell=1/2$. The real and imaginary parts of QNMs are shown, for both boundary conditions, in the first and second rows. We display imaginary part of QNMs in terms of real parts of QNMs in the third row. Notice that, we use solid lines with square (circle) dots to represent results for the first (second) boundary condition, the same color to stand for the same charge parameters $\tilde{Q}$ (left) and $q$ (right), and $\tilde{Q}=Q/Q_c$.}
\end{figure*}

In Fig.~\ref{chargeDiracNeffects}, a tower of QNMs with different overtone numbers $N$ are presented, by varying background (left panels) and field (right panels) charges. Here we take an intermediate BH ($r_+=1$) with $\ell=1/2$, as an illustration. In this figure we have used solid lines with square (circle) dots to represent QNMs with the first (second) boundary condition. The real (imaginary) part of QNMs is shown in the first (second) row. As one may observe, excited modes for both branches are approximately evenly spaced in $N$, varying either $Q$ or $q$. By increasing the background charge $Q$, for both sets of modes, the slope of real part decreases, while the slope of the imaginary part magnitude increases. Thus, the background charge affects QNMs more strongly for larger $N$. By increasing the field charge $q$, on the other hand, QNMs vary weakly for both set of modes, which implies that the field charge effect on QNMs for different $N$ is similar. A particularly interesting feature is shown in the third row, where we display the imaginary part in terms of the real part of the quasinormal frequency. QNMs (for excited modes) with first and second boundary conditions lie on the same line for different $N$. This means that although QNMs with two boundary conditions are different, they are similar in the sense that excited modes with the second boundary condition may be interpolated from the QNMs with the first boundary condition.

\section{Discussion and Final Remarks}
\label{discussion}
In this paper we have studied charged Dirac QNMs on RN-AdS BHs by imposing Robin type boundary conditions. These conditions follow from a generic physical principle: that the energy flux should vanish at the AdS boundary~\cite{PhysRevD.92.124006,Wang:2016dek}. To this end we first derived the charged Dirac equations by using the $\gamma$ matrices method and constructed the energy flux for charged Dirac fields. By requiring vanishing energy flux, we then obtained \textit{two} distinct sets of boundary conditions for charged Dirac fields. These conditions are similar to their counterparts for neutral Dirac fields, up to a gauge constant $C$ for the electrostatic potential. Without loss of generality, we fixed $C=0$, and performed both analytic and numeric calculations for charged Dirac QNMs under the two sets of boundary conditions, with various parameters, in particular exploring the effect of the electromagnetic charge.

For a small RN-AdS BH, for low frequency and small charge coupling $qQ$, we computed charged Dirac QNMs \textit{analytically} by using a matching method. The imaginary part of QNMs for charged Dirac fields with both boundary conditions was then shown to be always negative, implying the absence of superradiant instabilities. This contrasts with the case of charged bosonic fields which may trigger superradiant instabilities on RN-AdS BHs. This difference  between bosonic and fermionic fields can be understood both at the quantum and classical levels. At the quantum level, it is a consequence of Pauli's exclusion principle, which does not allow for more than one particle in each outgoing wave~\cite{Hawking:1974sw}. Therefore the scattered wave can not have a larger amplitude than the incident wave.  At the classical level, the area theorem does not apply to a fermion field since its energy-momentum tensor does not obey the weak energy condition~\cite{Hawking:1974sw}. Although superradiance is absent for charged Dirac fields, it seems that it is still possible to construct true bound states (or clouds) when a BH approaches extremality. For bosonic fields, an analogous sort of (marginal) clouds also exist~\cite{Degollado:2013eqa,Sampaio:2014swa}.

Beyond analytic approximations, we performed a numeric study for charged Dirac QNMs by using three different numeric methods. We mainly focused on the charge effect on the quasinormal frequencies. Let us briefly summarise the most relevant results.

We have chosen three different BH sizes ($r_+=0.1/1/100$) to explore charge effect on Dirac QNMs. For all three cases with both boundary conditions, by increasing the background charge $Q$ with fixed field charge $q=0$, the real part of the quasinormal frequency decreases. The only exceptional case is for $r_+=100$ with the second boundary condition, where the real part of QNMs first decreases and then increases as $Q$ increases. Turning on the field charge $q$, real part of QNMs with both boundary conditions for all three BH sizes, increases with $Q$ increases. The charge effect on the imaginary part of the QNMs depends on the BH size. By fixing $q=0$, the magnitude of the imaginary part of the QNMs, with both boundary conditions, first decreases and then increases for $r_+=0.1$, increases for $r_+=1$, and decreases for $r_+=100$, with increasing $Q$. By turning on the field charge $q$, the magnitude of the imaginary part,  with both boundary conditions, is similar for the cases of $r_+=0.1$ and $r_+=1$, i.e. they first decrease and then increase as $Q$ increases. For $r_+=100$, the magnitude of the imaginary part of the QNMs with both boundary conditions always decreases with increasing $Q$, for all field charges $q$ we considered.

The charge effect on Dirac QNMs in terms of the angular momentum quantum number $\ell$ was also explored. We observed that for both boundary conditions, the real part of the QNMs increases roughly linearly with increasing $\ell$; the magnitude of the imaginary part decreases weakly, for both cases, with varying background charge and field charge. Furthermore, we found that the charge effect on the QNMs  is more pronounced for the second boundary condition.

We also explored the charge effect on the Dirac QNMs with respect to the overtone number $N$. We found that, as for the neutral case, excited QNMs for both boundary conditions are approximately evenly spaced in $N$, for both varying $Q$ and $q$. By increasing the background charge $Q$, for both sets of modes, the slope of real part of QNMs decreases, while the slope of the magnitude of the imaginary part increases. By increasing the field charge $q$, we observed that the QNMs vary weakly for both set of modes. In particular, we found that the excited modes with the first and second boundary condition lie along the same straight line for different $N$. This means that although QNMs with the two boundary conditions are different, we may be able to obtain excited modes with the second boundary condition by simply interpolating excited QNMs with the first boundary condition.

The universality of the ``vanishing energy flux" principle has been verified by this work.  It can be applied, not only to bosonic fields  but also to fermionic fields in~\cite{Wang:2017fie}, and not only for neutral fields but also for charged fields, as in the present paper. To fully understand this principle for a Dirac field, the generalization of Dirac fields on rotating backgrounds is the next step. This analysis has not been performed even under the usual Dirichlet boundary conditions. Work along this direction is underway~\cite{rotatingDirac}.

{\bf Note added:}
After this paper was submitted for peer-reviewing, a paper appeared in the arXiv (1910.04181) with two criticisms on our previous work~\cite{Wang:2017fie}:
\\
$\bullet$\;\; firstly, 1910.04181 states that~\cite{Wang:2017fie} incorrectly claims  that \textit{the AdS/CFT correspondence no-source boundary conditions (...) do not yield Dirac field solutions with vanishing energy flux at  the  asymptotic  boundary  of  AdS}. A careful reading of~\cite{Wang:2017fie}, however, shows no such claim was ever made. \cite{Wang:2017fie} merely stated that these boundary conditions cannot be implemented in terms of the standard Teukolsky formalism variables $R_1$ or $R_2$, when solving the Dirac equation. The use of a combination of these variables, easily obtainable from \cite{Wang:2017fie}, allowed 1910.0418 to recover the QNM spectrum in~\cite{Wang:2017fie}, using the standard boundary conditions.
\\
$\bullet$\;\;  secondly,  1910.04181 states \cite{Wang:2017fie}  \textit{missed the negative spectrum}. Obviously, since the imaginary part of the negative and positive frequencies coincides, there is no new information in the negative part of the spectrum, which is the reason it was not mentioned in~\cite{Wang:2017fie}.

We appreciate the criticisms of 1910.04181. However, quoting Shakespeare, it is \textit{much ado about nothing}.

\bigskip

\noindent{\bf{\em Acknowledgements.}}
M. W would like to thank Q. Wang all the same, for her first inspiration and then desperation during the completion of this paper. This work is supported by the National Natural Science Foundation of China under Grant Nos. 11705054, 11881240252, and by the Hunan Provincial Natural Science Foundation of China under Grant No. 2018JJ3326. J.Jing's work is partially supported by the National Natural Science Foundation of China under Grant No. 11875025.  This work is supported by the Funda\c{c}\~ao para a Ci\^encia e a Tecnologia (FCT) project UID/MAT/04106/2019 (CIDMA) and project PTDC/FIS-OUT/28407/2017. This work has further been supported by  the  European  Union's  Horizon  2020  research  and  innovation  (RISE) programmes H2020-MSCA-RISE-2015 Grant No.~StronGrHEP-690904 and H2020-MSCA-RISE-2017 Grant No.~FunFiCO-777740. The authors would like to acknowledge networking support by the
COST Action CA16104. This research was also partially supported by the Munich Institute for Astro- and Particle Physics (MIAPP) of the DFG cluster of excellence ``Origin and Structure of the Universe", where part of this work was done there.

\bigskip

\appendix

\section{Functions and recurrence relations for Horowitz-Hubeny method}
\label{app1}
\noindent The functions in Eq.~\eqref{ChargedHHeq} are (we have set $L=1$)
\begin{align}
S(x)=&\aleph_1^2\;,\nonumber\\
T(x)=&\aleph_1\aleph_2\;,\nonumber\\
U(x)=&\;(\omega-qQx)^2-\varpi^2+iqQ\aleph_3+\dfrac{1}{2}i\omega\aleph_4-\lambda^2\aleph_5+\aleph_6\;,\nonumber
\\
\label{ChargedSeriesfunc}
\end{align}
where
\begin{align}
\aleph_1&=c_0+c_1x+c_2x^2-Q^2x^3\;,\nonumber\\
\aleph_2&=-2i\varpi-2x+3c_3x^2-4Q^2x^3\nonumber\\
\aleph_3&=-\dfrac{1}{2}\left(2+c_3x^3-2Q^2x^4\right)\;,\nonumber\\
\aleph_4&=-2x+3c_3x^2-4Q^2x^3\;,\nonumber\\
\aleph_5&=1+x^2-c_3x^3+Q^2x^4\;,\nonumber\\
\aleph_6&=\dfrac{1}{16}\left[8-24c_3x+4(1+12Q^2)x^2-20c_3x^3\right.\nonumber\\&\left.+(15c_3^2+40Q^2)x^4-48Q^2c_3x^5+32Q^4x^6\right]\;,
\end{align}
and
\begin{equation}
c_0=\dfrac{1}{x_+}\;,\;\;\;c_1=\dfrac{1}{x_+^2}\;,\;\;\;c_2=\dfrac{1+x_+^2}{x_+^3}\;,\;\;\;c_3=c_2+Q^2x_+\;.\nonumber
\end{equation}
\\
The recurrence relations between $a_j$ can be obtained by substituting the expansions of $\Phi_1$ into Eq.~\eqref{ChargedHHeq}.  We obtain
\begin{equation}
a_j=-\dfrac{1}{D_j}\sum_{n=1}^j[s_n(j-n)(j-n-1)+t_n(j-n)+u_n]a_{j-n}\;,\label{recurrrelation}
\end{equation}
where
\begin{equation}
D_j=s_0j(j-1)+t_0j\;.\nonumber
\end{equation}
\\

\bibliographystyle{h-physrev4}
\bibliography{DiracRNAdS} 

\end{document}